\documentclass[aps,prd,preprint,groupedaddress]{revtex4}

\usepackage{amsmath}
\usepackage{amssymb}
\usepackage{slashed}
\usepackage{latexsym}
\usepackage{dsfont}
\usepackage{epsfig}
\usepackage{graphicx}
\usepackage{psfrag}
\usepackage{subfigure}
\usepackage{placeins} 

\newcommand{\unitop}{\mathds{1}}
\newcommand{\bra}[1]{\left\langle #1 \right|}
\newcommand{\ket}[1]{\left| #1 \right\rangle}
\newcommand{\Tr}[1]{\mathrm{Tr} \left[ \, #1 \, \right]}
\newcommand{\Real}[1]{\mathrm{Re}\Big( \, #1 \, \Big)}
\newcommand{\Imag}[1]{\mathrm{Im}\Big( \, #1 \, \Big)}
\newcommand{\bea}{\begin{eqnarray}}
\newcommand{\eea}{\end{eqnarray}}
\newcommand{\be}{\begin{equation}}
\newcommand{\ee}{\end{equation}}
\newcommand{\wh}[1]{\widehat{#1}}
\newcommand{\nn}{\nonumber}
\newcommand{\onehalf}{\frac{1}{2}}

\newcommand{\eqnref}[1]{Eq.~(\ref{#1})}
\newcommand{\Eqnref}[1]{Eq.~(\ref{#1})}
\newcommand{\figref}[1]{Fig.~\ref{#1}}

\newcommand{\secref}[1]{Sec.~\ref{#1}}
\newcommand{\Secref}[1]{Sec.~\ref{#1}}
\newcommand{\appref}[1]{app.~\ref{#1}}

\def\BA{\begin{eqnarray}}
\def\BE{\begin{equation}}
\def\EA{\end{eqnarray}}
\def\EE{\end{equation}}

\def\gtsim{\lower-0.45ex\hbox{$>$}\kern-0.77em\lower0.55ex\hbox{$\sim$}}
\def\ltsim{\lower-0.45ex\hbox{$<$}\kern-0.77em\lower0.55ex\hbox{$\sim$}}

\newcommand{\link}{\mbox{
\begin{picture}(0.2, 1.1)
\thicklines
\put(0.05,0.1){\vector(0,1){1}}
\put(-0.1,-0.4){\scriptsize \( \vec{y} \) }
\end{picture}}}
\newcommand{\staplel}{\mbox{
\begin{picture}(1.1, 1.1)
\thicklines
\put(1,0.1){\vector(-1,0){1}}
\put(0,0.1){\vector(0,1){1}}
\put(0,1.1){\vector(1,0){1}}
\put(-0.6,-0.6){\scriptsize \( \vec{y}-\vec{e}_- \) }
\end{picture}}}

\newcommand{\stapler}{\mbox{
\begin{picture}(1.7, .1)
\thicklines
\put(0,0.1){\vector(1,0){1}}
\put(1,0.1){\vector(0,1){1}}
\put(1,1.1){\vector(-1,0){1}}
\put(0.4,-0.6){\scriptsize \( \vec{y} +\vec{e}_- \) }
\end{picture}}}

\newcommand{\staplefp}{\mbox{
\begin{picture}(1.7, .1)
\thicklines
\put(0,0.1){\vector(0,1){1}}
\put(0,1.1){\vector(1,0){1}}
\put(1,1.1){\vector(0,-1){1}}
\put(1,0.1){\vector(-1,0){0.8}}
\put(0.2,0.1){\vector(0,1){1}}
\put(0.2,1.1){\line(0,1){0.2}}
\put(0.4,-0.6){\scriptsize \( \vec{y} +\vec{e}_- \) }
\end{picture}}}

\newcommand{\staplebp}{\mbox{
\begin{picture}(1.1, 1.1)
\thicklines
\put(1,0.1){\vector(0,1){1}}
\put(1,1.1){\vector(-1,0){1}}
\put(0,1.1){\vector(0,-1){1}}
\put(0,0.1){\vector(1,0){0.8}}
\put(0.8,0.1){\vector(0,1){1}}
\put(0.8,1.1){\line(0,1){0.2}}
\put(-0.6,-0.6){\scriptsize \( \vec{y}-\vec{e}_- \) }
\end{picture}}}

\begin{document}

\title{Gluon Structure Function of a Color Dipole \\
in the Light-Cone Limit of Lattice QCD }

\author{D. Gr\"unewald}
\email[]{d.gruenewald@tphys.uni-heidelberg.de}
\affiliation{Institut f\"ur Theoretische Physik, Universit\"at Heidelberg,
Germany}
\author{E.-M. Ilgenfritz}
\email[]{ilgenfri@physik.hu-berlin.de}
\affiliation{Institut f\"ur Physik, Humboldt-Universit\"at zu
Berlin, Germany}
\affiliation{Institut f\"ur Theoretische Physik, Universit\"at Heidelberg,
Germany}
\author{H.J. Pirner}
\email[]{pirner@tphys.uni-heidelberg.de}
\affiliation{Institut f\"ur Theoretische Physik, Universit\"at Heidelberg,
Germany}
\affiliation{Max-Planck-Institut f\"ur Kernphysik Heidelberg, Germany}

\date{\today}

\begin{abstract}
We calculate the gluon structure function of a color dipole in 
near-light-cone SU(2) lattice QCD as a function of $x_B$. 
The quark and antiquark are external non-dynamical degrees of freedom
which act as sources of the gluon string configuration defining the dipole. 
We compute the color dipole matrix element 
of transversal chromo-electric and 
chromo-magnetic field operators separated along a direction close to the light cone,
the Fourier transform of which is the gluon structure function. As vacuum state in the pure
glue sector, we use 
a variational ground state of the near-light-cone Hamiltonian.
We derive a recursion relation for the gluon structure function
on the lattice similar to the perturbative DGLAP 
equation. It depends on the number of transversal links 
assembling the Schwinger string of the dipole. Fixing the mean momentum fraction of the
gluons to the "experimental value" in a proton, we compare our gluon structure 
function for a dipole state with four links with the NLO \emph{MRST} 2002
and the \emph{CTEQAB-0} parameterizations at $Q^2=1.5\,\mathrm{GeV}^2$. Within
the systematic uncertainty we 
find rather good agreement. We also discuss the low $x_B$ behavior of the gluon structure
function in our model calculation. 
\end{abstract}


\maketitle

\section{Introduction}

First support for QCD as the theory of strong interactions has come from 
deep inelastic scattering. The structure of the proton unfolds itself 
in terms of partons which interact only weakly due to asymptotic freedom.
The deduced structure functions representing the proton constituents
are Fourier transforms of quark/gluon operators separated by light like 
distances. Theoretical calculations of scaling violations by the DGLAP 
equation \cite{Gribov:1972ri,Altarelli:1977zs,Dokshitzer:1977sg}
have strongly contributed to the understanding of deep inelastic scattering.
A current QCD analysis of experiments is given e.g. in 
Refs.~\cite{Chekanov:2002pv,Martin:2001es,Pumplin:2005rh}. For small $x_B$, there appear 
-- in addition to terms from DGLAP-evolution  -- other BFKL-contributions 
like $\alpha_s(Q^2) \log (1/x_B)$ \cite{Fadin:1975cb,Balitsky:1978ic,Fadin:1998py} 
which need special care. Also, 
it is needless to stress that in a perturbative 
framework  structure functions themselves cannot be calculated from first principles. 
Euclidean lattice simulations use the operator product expansion to get information 
about quark structure functions. In this way, the lowest moments of the meson and 
nucleon structure functions have been evaluated \cite{Best:1997qp,Gockeler:2004wp,Negele:2004iu}. 
Nowadays, this method has been generalized to non-forward matrix elements (generalized
parton distribution functions) \cite{Hagler:2003jd}. Recently, loop-loop correlation 
functions of tilted Wegner-Wilson loops have been computed on a Euclidean 
lattice \cite{Giordano:2008ua} which can be related to the gluon distribution 
function \cite{Nikolaev:1990ja,GolecBiernat:1999qd,Shoshi:2002fq} of a color dipole 
or a hadron. 

Parallel to these investigations, the light cone lattice community has pursued
\cite{Dalley:2003aj,Vary:2009gt,Naus:1997zg} the idea of a different formulation of QCD on 
or near the light cone. The hope is that a theoretical framework based on constituents 
moving along the light cone will be simple, rather closely following the course of the 
experimental discovery of quarks. Of course, the light cone approach has to attempt to 
incorporate the non-perturbative QCD vacuum, which is hard to achieve in the framework
of a Fock representation of free fields acting on a trivial vacuum. Also rewriting a 
spatially quantised theory into a theory quantised on a light like surface may present 
problems. 

Therefore, in a recent paper \cite{Grunewald:2007cy} we have developed a 
near-light-cone (nlc) approach in which we can combine the advantages of the lattice 
formulation with the advantages of light cone simplifications. In this reference, we 
have constructed a ground state wave functional of the near-light-cone Hamiltonian which, 
in the light cone limit, becomes simpler than the equal time ground state   
in the similar strong coupling approximation. Here, in a first application we use this 
variationally optimized ground state wave functional to determine the gluon distribution 
function of a color dipole. A color dipole is a 
system consisting of a static quark and antiquark pair connected by a Schwinger string.  
In our simplified dipole picture, we use the average momentum fraction carried by the 
gluons as extracted from phenomenological analyses as an input. We then predict the shape 
of the gluon structure function as a function of the transverse size of the dipole.
One should remark that most dipole calculations for the gluon distribution are done in 
a reference system where the hadron under consideration is at rest. In our calculation 
the hadron is attached to the fast moving frame. This justifies the application of 
near-light-cone dynamics for its constituents. 

The outline of the paper is as follows: In \secref{sec:ContGluonDist} we review the original 
definition of the gluon structure function on the light cone and some of its properties. 
Next, we define the gluon structure function near the light cone such that its original 
definition is recovered in the light cone limit. In \secref{sec:NLCHamiltonian}, we 
recapitulate properties of the near-light-cone (nlc) lattice formulation. This leads us
to the variationally optimized ground state wave functional. In \secref{sec:DipoleModel} 
the model of the hadron as a color dipole state is outlined. 
We define the lattice counterpart of the nlc correlation function in \secref{sec:NLCCorrLat}.
In \secref{sec:GluonDistSingleLink}, we discuss the lattice computation of the gluon structure 
function for a one-link dipole, which yields the building block for the computation of hadronic 
gluon distribution functions.  
\Secref{sec:HadronicGluonDist} contains our results and their interpretations. 
In \secref{sec:Summary}
we formulate our conclusions and discuss possible improvements.  

\section{Gluon distribution function on and close to the light cone}
\label{sec:ContGluonDist}
First, we review the original definition and some properties of the gluon distribution function. 
In deep inelastic scattering the hadronic target is probed on the light cone, i.e. at equal light
cone time $x^+=0$. Here, $x^+=(x^0+x^3)/\sqrt{2}$ and $x^-=(x^0-x^3)/\sqrt{2}$ are the ordinary 
light cone temporal and the light cone longitudinal coordinate. 
In light cone quantisation one quantises on a light like hypersurface defined by $x^+=0$.
On this hypersurface, the entire scattering process is static. Therefore, there is no need to 
evolve the hadronic wave function in light cone time during scattering.

The internal structure of the hadronic target is encoded in parton distribution 
functions. For example, the gluon distribution function $g(x_B)$ represents the probability 
that a gluon carries the longitudinal momentum fraction $x_B$ of the fast moving hadronic 
target \cite{Collins:1981uw}. In light cone coordinates, it is given by the 
Fourier transform of the 
matrix element of the two-point operator
$G(z^-,\vec{z}_\perp\,;0,\vec{z}_\perp)$ 
of longitudinally separated gluon field strength operators in a hadron state
$|h(p_-,\vec{0}_\perp)\rangle$:
\be
g(x_B)=\frac{1}{x_B}\,\frac{1}{2\,\pi}\int_{-\infty}^{\infty} dz^-\,d^2\vec{z}_\perp
\,e^{-\mathrm{i}\,x_B\,p_-\,z^-}\,\frac{1}{p_-}\bra{h(p_-,\vec{0}_\perp)}G(z^-,\vec{z}_\perp\,;0,\vec{z}_\perp)
\ket{h(p_-,{\vec 0}_\perp)}_c \,. 
\label{eqn:DefGluonDistr}
\ee
The notation $|h(p_-,\vec{0}_\perp)\rangle$ emphasizes that the hadron is localized with its 
center of mass in transversal configuration space at 
$\vec{b}_\perp=\vec{0}_\perp$ and carries longitudinal momentum $p_-$.
The momentum $x_B$ is normalized 
relative to the total momentum $p_-$ of the entire hadron. 
The index ``c'' indicates that the connected matrix element is taken, i.e. the product of 
the vacuum matrix element with the normalization of the hadronic state is subtracted: 
\bea
\lefteqn{
\hspace{-4.2cm}
\bra{h(p_-,\vec{0}_\perp)}\,G(z^-,\vec{z}_\perp\,;\,0,\vec{z}_\perp)\,\ket{h(p_-,\vec{0}_\perp)}_c}\nn\\
&=&
\phantom{-}\bra{h(p_-,\vec{0}_\perp)}\,G(z^-,\vec{z}_\perp\,;\,0,\vec{z}_\perp)\,\ket{h(p_-,\vec{0}_\perp)}\nn\\
& &
-\Big\langle\Omega\Big|\,G(z^-,\vec{z}_\perp\,;\,0,\vec{z}_\perp)\,\Big|\Omega\Big\rangle\,\left\langle
  h(p_-,\vec{0}_\perp)\right.\ket{h(p_-,\vec{0}_\perp)}\,,
\nn\\
\bra{h(p_-,\vec{0}_\perp)}\left.h(p_-,\vec{0}_\perp)\right\rangle &=&2\,p_-\,L_- \, .  
\label{eqn:DefGluonDist}
\eea
Here, $\ket{\Omega}$ denotes the vacuum state. 
$L_-$ is the spatial extension along the longitudinal direction of the normalization box.
The point split operator $G(z^-,\vec{z}_\perp\,;\,0,\vec{z}_\perp)$ corresponding to the wanted correlation function is given by \cite{Collins:1981uw} 
\be
G(z^-,\vec{z}_\perp\,;\,0,\vec{z}_\perp)=\sum\limits_{k=1}^2 F_{-k}^a(z^-,\vec{z}_\perp)\,S_{a b}^A(z^-,\vec{z}_\perp\,;\,0,\vec{z}_\perp)\,F_{-k}^b(0,\vec{z}_\perp)
\label{eqn:LFCorrFunc}
\ee
with 
\be
S_{ab}^A(z^-,\vec{z}_\perp\,;\,0,\vec{z}_\perp)=
\Big[
\mathcal{P}\exp\Big\{\mathrm{i}\,g\,
\int_{0}^{z^-}dx^-\,A_{-}^c(x^-,\vec{z}_\perp)\,\lambda^c_{adj} \Big\}
\Big]_{ab}\,.
\ee
The Schwinger string in the adjoint representation 
$S_{ab}^A(z^-,\vec{z}_\perp\,;\,0,\vec{z}_\perp)$ connects the gluon field strength 
operator at the point $\vec{z}_\perp$ in the $z^{-}=0$ plane, $F_{-k}^b(0,\vec{z}_\perp)$, 
with the longitudinally separated field strength operator $F_{-k}^a(z^-,\vec{z}_\perp)$ along a 
light like path.  In the usual light cone quantisation approach, one uses the so called
light cone gauge $A_-= A^+=0$ for quantisation. This sets the Schwinger string along the light cone equal to one. 
The importance of the Schwinger strings along the light cone is visualized e.g. in
the loop-loop correlation model where hadron-hadron scattering cross-sections
are calculated from Wegner-Wilson loop correlation functions \cite{Shoshi:2002rd,Shoshi:2002in}. 
The eikonal phases arising from the strings along the $x^-$ direction also describe the 
so-called ``final state'' interaction effects which distinguish structure functions from 
parton probabilities \cite{Brodsky:2002ue}.

The gluon distribution function defined in that way obeys a momentum sum rule, i.e.
the average momentum fraction of the hadron carried by the gluons is related to
the first moment of the structure function. The integral over $x_B$ can be reformulated 
as an integral over gluon momenta $p_-^g$, which yields the matrix element
of the gluonic two-point operator $G$ taken at $z^-=0$: 
\bea
\left\langle x_B \right\rangle 
&=& \int_0^1 dx_B \, x_B \, g(x_B)\nn\\
&=&\frac{1}{2\,\pi} \int_{-\infty}^\infty dz^-\,d^2\vec{z}_\perp \int_0^\infty \frac{dp_-^g}{p_-^2} \,e^{-\mathrm{i}\,p_-^g\,z^-}\,\bra{h(p_-,\vec{0}_\perp)}G(z^-,\vec{z}_\perp\,;\,0,\vec{z}_\perp)\ket{h(p_-,\vec{0}_\perp)}_c \nn\\
&=&
\int_{-\infty}^\infty dz^-\,d^2\vec{z}_\perp \frac{1}{2\,p_-^2}\delta(z^-)\,\bra{h(p_-,\vec{0})}G(z^-,\vec{z}_\perp\,;\,0,\vec{z}_\perp)\ket{h(p_-,\vec{0}_\perp)}_c  \; ,
\label{eqn:AvgxB}
\eea
which coincides with the longitudinal light cone momentum density operator 
$\mathcal{P}_-^{lc}$ 
\be
\left. G(z^-,\vec{z}_\perp\,;\,0,\vec{z}_\perp)\right|_{z^-=0}=\mathcal{P}_-^{lc}(0,\vec{z}_\perp)
=\sum_{k=1}^2F_{-k}^a(0,\vec{z}_\perp)\,F_{-k}^a(0,\vec{z}_\perp)\,.
\ee
Hence, the average gluon fractional momentum is given by 
\be
\left\langle x_B \right\rangle =
 \frac{1}{2\,p_-^2}  \int d^2\vec{z}_\perp\,\bra{h(p_-,\vec{0}_\perp)}\mathcal{P}_-^{lc}(0,\vec{z}_\perp)\ket{h(p_-,\vec{0}_\perp)}\,.
\label{eqn:AvgGM}
\ee
The subscript $c$ could be dropped since the 
disconnected part of the matrix element vanishes.
This is the case because  
the expectation value of
the longitudinal momentum density of the vacuum state vanishes, see \secref{sec:NLCCorrLat}.
\eqnref{eqn:AvgGM} yields the average fractional gluon momentum
with the normalization of the hadronic state given in \eqnref{eqn:DefGluonDist}. 

To compute the gluon distribution function non-perturbatively on the lattice, we 
shall use the light cone limit of ``near to the light cone'' (nlc) quantisation instead.
Here, nlc refers to near-light-cone coordinates \cite{Prokhvatilov:1989eq,lenz} which 
have been introduced to implement light front quantisation as a limit
of equal time quantisation. The nlc transverse and
longitudinal coordinates $\vec{x}_\perp$ and $x^-$ are defined in a similar way
as usual light cone coordinates. The definition of the temporal
nlc coordinate $x^+$ however contains an additional external parameter
$\eta$ which parameterizes a rotation in the $x^0-x^3$ plane
not included in the Lorentz group and which allows for a smooth
interpolation between equal time quantisation
($\eta=1$ , $x^+= x^0$) and light cone quantisation
($\eta=0$ , $x^+=1/2\,( x^0+x^3)$). 
\bea
x^+&=&\frac{1}{2}\Bigl[\left(1+\eta^2\right)x^{0}+ \left(1-\eta^2\right)x^{3}\Bigr] \nonumber\\
x^-&=&\phantom{\frac{1}{2}}\Bigl[x^{0}-x^{3}\Bigr] \; .
\eea
Note that the $\eta \to 0$ limit can be interpreted as the infinite momentum frame limit 
in which the partons of the color dipole move with infinite momentum. Quantisation in 
terms of near-light-cone coordinates in contrast to ordinary light cone quantisation 
has the advantage that no quantum constraint equations have to be solved. This makes a 
lattice treatment feasible, at least in the Hamiltonian formulation. 

In a Hamiltonian nlc theory obeying the standard $A_+=0$ gauge, the operator of the 
longitudinal momentum density in the pure gauge sector can be obtained from the energy 
momentum tensor by expressing the temporal components of the field strength tensor in 
terms of the chromo-electric field operators similar to the usual Legendre transformation 
from the Yang-Mills Lagrange density to the Hamiltonian density. In the $A_+=0$ gauge, 
the chromo-electric field operators are given by the functional derivatives of the 
Lagrange density with respect to the space-time components of the field strength tensor.     
Hence, the operator of the longitudinal momentum density is given by \cite{Grunewald:2007cy} 
\be
\mathcal{P}_-(\vec{z})=  
\onehalf \sum\limits_{k=1}^2\Big( \Pi_k^a(\vec{z})\,F_{-k}^a(\vec{z})+F_{-k}^a(\vec{z})\,\Pi_k^a(\vec{z})\Big)\,.
\label{eqn:DefLongMomDens}
\ee
Here, $\Pi_i^a(\vec{z})$ is the chromo-electric field operator which is canonically conjugate
to the gauge field $A_j^b(\vec{y})$, i.e.
\be
\left[\Pi_i^a(\vec{z}),A_j^b(\vec{y}) \right]= 
-\mathrm{i}\,\delta^{(3)}({\vec{z}-\vec{y}})\,\delta_{i,j}\,\delta^{a,b}~~,~~i,j=1,2,- \,.
\label{eqn:cancommrel}
\ee  
The longitudinal momentum density in \eqnref{eqn:DefLongMomDens} is symmetrized 
in order to render it hermitean. This combination of transversal chromo-electric field 
operators and chromo-magnetic field operators is quite natural because it resembles the 
Poynting vector in ordinary electrodynamics representing the momentum density of the 
electromagnetic field in longitudinal direction. In the light cone limit, the transversal 
chromo-magnetic field strength operators become equal to the corresponding chromo-electric 
field strength operators due to the constraint equation which emerges in the light cone limit.
 
In order to have the same momentum sum rule in near-light-cone coordinates as one has in 
light cone coordinates, we define the operator corresponding to the near-light-cone 
correlation function as a point-split generalization of the longitudinal momentum density 
given in \eqnref{eqn:DefLongMomDens}
\bea
G(z^-,\vec{z}_\perp;z^\prime{}^-,\vec{z}_\perp)&=&\frac{1}{4}\,\frac{1}{L_-}\sum\limits_{k=1}^2 \Big(\Pi_k^a(z^-,\vec{z}_\perp)\,S_{a b}^A(z^-,\vec{z}_\perp\,;z^\prime{}^-\,\vec{z}_\perp)\,F_{-k}^b(z^\prime{}^-,\vec{z}_\perp)\nn\\
& &~~~~~~~~~~~+\Pi_k^a(z^\prime{}^-,\vec{z}_\perp)\,S_{a b}^A(z^\prime{}^-,\vec{z}_\perp\,;z^-\,\vec{z}_\perp)\,F_{-k}^b(z^-,\vec{z}_\perp)+h.c.\Big)\,. 
\label{eqn:nlcCorrFunc}
\eea
We have symmetrized also this operator with respect to an interchange 
$z^-\leftrightarrow z^\prime{}^-$ and with respect to the ordering of the transversal 
chromo-magnetic and -electric field operators. In the light cone limit, this operator 
reproduces the definition \eqnref{eqn:DefGluonDistr}. 
Note that in our nlc Hamiltonian approach the gauge fields $A_-$ are fully dynamical 
gauge fields, only the gauge choice $A_+=0$ has been implemented.

For later convenience, we use translation invariance of the expectation value in order to 
introduce an additional integration over the longitudinal coordinate $z^\prime{}^-$.
Using the above operator \eqnref{eqn:nlcCorrFunc} and the normalization of the hadronic 
state \eqnref{eqn:DefGluonDist}, the gluon distribution function in nlc coordinates is given by
\be
g(x_B)=\lim_{\eta \to 0} \,\frac{1}{2\,\pi} \frac{1}{x_B}\,\int_{-\infty}^{\infty} dz^-\,dz^\prime{}^-\,d^2z_\perp\,e^{-\mathrm{i}\,x_B\,p_-(z^--z^\prime{}^-)}
\,\widetilde{g}(z^-,\vec{z}_\perp;z^\prime{}^-,\vec{z}_\perp)
\label{eqn:nlcDefGluonDistr}
\ee
with the abbreviation
\be
\widetilde{g}(z^-,\vec{z}_\perp;z^\prime{}^-,\vec{z}_\perp)= 
2\, \frac{
\bra{h(p_-,\vec{0}_\perp)}G(z^-,\vec{z}_\perp;z^\prime{}^-,\vec{z}_\perp) \ket{h(p_-,\vec{0}_\perp)}_c}
{\left\langle h(p_-,\vec{0}_\perp)\right.\left| h(p_-,\vec{0}_\perp)\right\rangle}
\,.
\label{eqn:nlcDefFTGluonDistr}
\ee
Note that the longitudinal nlc momentum $p_-$ of the target state is restricted 
to positive values in the light cone limit. If $p_-$ is expressed in terms of 
the ordinary Minkowski space momentum $p_3$, one obtains for an on-shell
particle like the target hadron the following expression for the longitudinal momentum
in the nlc frame:
\be
p_-=\left\{
\begin{array}{lcl}
-\eta^2\,p_3+\onehalf(1-\eta^2)\frac{m_\perp^2}{p_3} & , & p_3\geq0 \\
-p_3-\onehalf(1-\eta^2)\frac{m_\perp^2}{p_3} & , & p_3<0
\end{array}
\right.
+\mathcal{O}\left(\frac{m_\perp^4}{|p_3|^3} \right)
~~\stackrel{\forall\,|p_3|\gg m_\perp}{\Rightarrow} ~\lim_{\eta \to 0} p_-\in[0,\infty]\,. 
\label{eqn:posLatMom}
\ee    
Here, $m_\perp^2=\vec{p}_\perp^2+m^2$ is the transversal mass.

\section{Near-light-cone lattice Hamiltonian}
\label{sec:NLCHamiltonian}

In our previous work~\cite{Grunewald:2007cy} we have regularized $SU(2)$ gauge theory for 
our purposes by introducing a spatial nlc lattice. The size is 
$L_ -L_{\bot}^2=N_-a_- N_{\bot}^2a_{\bot}^2$, where $N_-,N_{\bot}$ are the number 
of lattice sites along the light like direction $x^-$ and the transversal directions $x_{k}$. 
The lattice spacings in these directions are $a_-$ and $a_{\bot}$. In the following, 
all spatial and momentum variables are assumed to be made dimensionless lattice quantities by
multiplication with the appropriate powers of the lattice spacings in longitudinal
and transversal directions. The gauge field degrees of freedom on the lattice are given by the 
gluon link operators $U_j(\vec{x})$ 
\be
U_j(\vec{x})=\mathcal{P}\exp\left(\mathrm{i}\,
g\int_0^1 dy\,a_j\,A_j(\vec{x}+y\wh{e}_j) \right)\,.
\ee
The ordering symbol $\mathcal{P}$ orders the products of gluon fields
$A_j(\vec{x}+y\wh{e}_j)$ from left ($y=0$) to right $y=1$).
In Hamiltonian theory we have two transversal gauge fields $A_j,(j=1,2)$
and one longitudinal gauge field  $A_-,(j=-)$. The $A_+$ component of the gauge field 
is set equal to zero in the Hamiltonian approach. As a result one has the Gauss-law
constraint which restricts the entire Hilbert space to the physical sector of 
gauge invariant states. The gluon dynamics is determined by the effective nlc lattice 
Hamiltonian which has been derived in Ref.~\cite{Grunewald:2007cy}. It represents 
the gluon energy density on the lattice. The QCD coupling constant enters as 
$\lambda=4/g^4$ in the $SU(2)$ Hamiltonian
\bea
\mathcal{H}_{\mathrm{eff,lat}}&=& \frac{1}{N_-N_{\bot}^2}\frac{1}
{a_{\bot}^4} \frac{2}{\sqrt{\lambda}}
\sum\limits_{\vec{x}}\left\{
    \frac{1}{2}~\sum\limits_{a}~\Pi_-^a(\vec{x})^2~
    +\onehalf~\lambda~\Tr{\unitop-\Real{{U}_{12}(\vec{x})}}
    \right.  \nn\\
& &  \left.
    ~+\sum\limits_{k,a}~\frac{1}{2}~\frac{1}{\tilde{\eta}^2}~ 
      \Bigg[~\Pi_k^a(\vec{x})^2+
      \lambda~\Biggl(\Tr{\frac{\sigma_a}{2}~\Imag{U_{-k}(\vec{x})}}\Biggr)^2
      \Bigg] 
    \right\}\; .
\label{eqn:EffectiveLatHamiltonian}    
\eea
The $2\times 2$ Pauli matrices $\sigma^a/2$ are the generators of the fundamental 
representation of the group $SU(2)$. The Hamiltonian depends on the gluon link 
operators through the "real" and "imaginary" parts of the plaquette operators 
$U_{12},U_{-k}$, which are obviously 
\be
\mathrm{Re}\left(U_{ij}\right)\equiv\frac{U_{ij}+U_{ij}^{\dagger}}{2}\equiv\frac{1}{2}~\Tr{U_{ij}} ~~ ,~~  
  \mathrm{Im}\left(U_{ij}\right)\equiv\frac{U_{ij}-U_{ij}^{\dagger}}{2\mathrm{i}} \,,
\label{def:ReIm}
\ee
for
\be
U_{ij}(\vec{x})=U_{i}(\vec{x})U_{j}(\vec{x}+\wh{e}_i)U_{i}^{\dagger}(\vec{x}+\wh{e}_j)U_{j}^{\dagger}(\vec{x})~~,~~\,i,j=1,2,-\,.
\label{eqn:DefPlaq}
\ee
We use $\mathrm{Im}\left[U_{ij}\right]$ as an abbreviation for the 
antihermitean part of the plaquette (which is traceless for $SU(2)$) 
$U_{ij}$ and $\mathrm{Re}\left[U_{ij}\right]$ (which is a multiple of 
the unit matrix for $SU(2)$) to represent its trace.
The Hamiltonian also contains the dimensionless lattice chromo-electric field strength 
operators $\Pi_i^a(\vec{x})$, which are canonically conjugate to the links. They 
obey the lattice commutation relations which follow directly from the continuum commutation 
relations in \eqnref{eqn:cancommrel},
\be
\left[\Pi_i^a(\vec{x}),U_j(\vec{y}) \right]= 
\frac{\sigma^a}{2}\,U_i(\vec{x})\,\delta_{\vec{x},\vec{y}}\,\delta_{i,j}\,.
\label{eqn:cancommrellat}
\ee
The constant $\tilde{\eta}$ is the product of the near-light-cone parameter $\eta$
and the anisotropy parameter $\xi=a_-/a_\bot$, 
\be
\tilde{\eta} = \eta \cdot \xi \, .
\ee 
If one chooses $\eta=1$ and varies $\xi$, one simulates an anisotropic equal time theory
with a ratio  $\xi=a_-/a_{\bot}$ of lattice constants $a_-$ and $a_{\bot}$ in longitudinal 
and transverse directions. In the limit $\xi \to 0$ one ends up with a system,
which is contracted in the longitudinal direction. Verlinde and Verlinde \cite{Verlinde:1993te} 
and Arefeva \cite{Arefeva:1993hi} have advocated such a lattice to describe high energy 
scattering. A contracted longitudinal system means that even the minimal momenta become 
high in longitudinal direction which is a promising starting point for high energy scattering. 
It is obvious that this limit leads to the same physics as the light cone limit with equal 
lattice constants in longitudinal and transverse directions while $\eta \to 0$.
In both cases the near-light-cone Hamiltonian is dominated by the terms proportional 
to $1/\tilde{\eta}^2$ involving transverse chromo-electric and chromo-magnetic fields.

In Ref.~\cite{Grunewald:2007cy} we have determined a variational gluonic ground state wave 
functional $\ket{\Psi_0}$ which consists of a product of single-plaquette wave functionals 
with two variational parameters $\rho$ and $\delta$ 
\bea
\ket{\Psi_0}&=&\Psi_0[U]\,\ket{0}=\sqrt{N_{\Psi}}\,e^{f[U]}\,\ket{0} \, , \nonumber \\
f[U]                          &=& 
\sum\limits_{\vec{x}} \left\{\sum\limits_{k=1}^2 
                 \rho_0(\lambda,\tilde{\eta})\,\Tr{\Real{U_{-k}(\vec{x})}}
                +\delta_0(\lambda,\tilde{\eta})\,\Tr{\Real{U_{12}(\vec{x})}}
               \right\} \, .
\label{eqn:VariationlGSWF}
\eea 
$N_{\Psi}$ is a normalization factor.
Here, the state $\ket{0}$ represents the trivial ground state
which is annihilated by the field momenta $\Pi_k^a(\vec{x})$ conjugate to the links,
\be
\Pi_k^a(\vec{x})\,\ket{0}=0\,\,\mathrm{and}\,\,\bra{0}\,\Pi_k^a(\vec{x})=0\,\,~\mathrm{for~all}~\,\vec{x},k,a\,.
\label{Def:StrongCouplingGroundState}
\ee
This ground state wave functional is similar to the ground state wave functional used 
in the strong coupling limit of equal-time quantised lattice gauge theory
\cite{Chin:1985ua}. However, it takes into account the anisotropy 
of the gauge dynamics 
in the purely transversal and the transversal-longitudinal planes. As in the equal time 
case, keeping the wave functional restricted to the one-plaquette form does not allow
to perform a continuous approach to the continuum limit. Further possible improvements 
are discussed in Ref. \cite{gruenewald_thesis}. We have optimized this ansatz with respect 
to the expectation value of the Hamiltonian over a large region in coupling space.
We are in the position to extrapolate the parameters $\rho_0,\delta_0$ to the light cone 
$\tilde{\eta} \to 0$. This limit yields the following remaining dependence on $\lambda$
\bea \rho_0(\lambda,0)&=&
\left(0.65-\frac{0.87}{\lambda}+\frac{1.65}{\lambda^2}\right)\sqrt{\lambda}
\, ,\nn\\ \delta_0(\lambda,0)&=&
\left(0.05+\frac{0.04}{\lambda}-\frac{1.39}{\lambda^2}\right)\sqrt{\lambda}
\; .
\label{eqn:OptParam}
\eea
\noindent
A typical value of $\lambda$ used in the subsequent calculations is $\lambda=10$,
 for which one obtains
\bea
 \rho_0(10,0)&=& 1.83\nn\\
 \delta_0(10,0)&=& 0.13\,.
 \label{eqn:OptParamLam10}
\eea 
Note that the ground state \eqnref{eqn:VariationlGSWF} is an approximation
found for the fully interacting effective Hamiltonian and does not rely on 
any truncated
Fock space expansion around the perturbative vacuum. 

In the light cone limit $\tilde{\eta} \to 0$ of the ground
state wave functional \eqnref{eqn:VariationlGSWF}, under averaging with the weight 
$|\Psi_0[U]|^2$, the
behavior of the gauge fields in the transverse $(1,2)$ 
plane is strongly coupled 
as shown by the parameter $\delta_0(\lambda,0)$ becoming small 
(cf. \eqnref{eqn:OptParamLam10}).
As a consequence of this, in the limit $\delta_0(\lambda,0) \to 0$, a strong coupling 
approximation turns out to be valid also in the transversal-longitudinal plane 
even for values of $\lambda$ which are far from the region $\lambda<<1$, say $\lambda=10$, 
as proven by actual Monte Carlo sampling of the squared ground state wave 
functional \eqnref{eqn:VariationlGSWF} \cite{gruenewald_thesis}. In the light
cone limit $\tilde{\eta} \to 0$ the gauge dynamics in each of
the hyperplanes $(-,1)$ and $(-,2)$ becomes 
two dimensional (and decoupled).
In two dimensions with free boundary conditions, moreover, the strong coupling
approximation is exact. 
\begin{figure}
\centering
\includegraphics[width=0.6\textwidth]{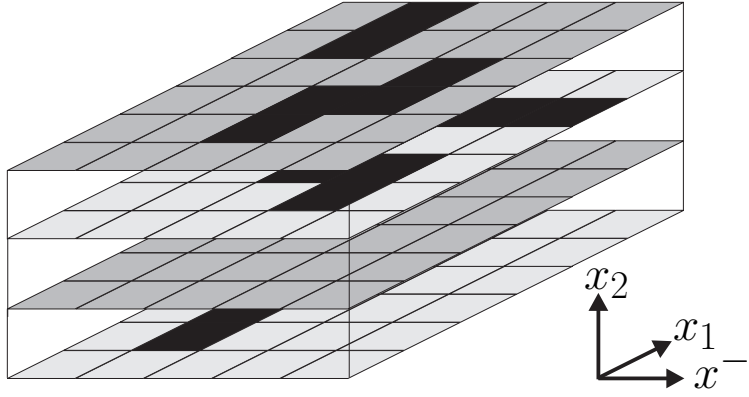}
\caption{\label{fig:Decoupling} In the light cone limit of the 
nlc Hamiltonian, the ground state wave functional decouples the dynamics of the $(x^-,x_1)$- and
 $(x^-,x_2)$-planes from each other. 
Thus, one obtains many decoupled two-dimensional gauge theories.
The planar structures (Wegner-Wilson loops) shown in black sketch vacuum 
fluctuations described
by the ground state wave functional inside of the $(x^-,x^1)$-planes. }
\end{figure}
These are the reasons why Hamiltonian gluon dynamics on the light cone 
is considerably simplified compared with 
equal time Hamiltonian QCD. We have the following standard area law 
behavior for Wegner-Wilson loops $W$ in the $(x^-,1)$ and $(x^-,2)$ planes, namely for
\be
W(0,\vec{0}_\perp\,;\,z{}^-,\vec{d}_\perp) = 
S_-(0,\vec{0}_\perp\,;\,z^-,\vec{0}_\perp)\,S_\perp(z^-,\vec{0}_\perp\,;\,z^-,\vec{d}_\perp)
  S_-(z^-,\vec{d}_\perp\,;\,0,\vec{d}_\perp)\,S_\perp(0,\vec{d}_\perp\,;\,0,\vec{0}_\perp) 
\ee
expressible through the plaquettes
\be
\bra{\Psi_0} \onehalf\, \Tr{W(0,\vec{0}_\perp\,;\,z{}^-,\vec{d}_\perp)} \ket{\Psi_0}
=\left(\bra{\Psi_0} \onehalf \Tr{U_{-k}}\ket{\Psi_0}\right)^{d_\perp\,|z{}^-|} \; .
\label{eqn:StrongCouplingApprox}
\ee
The physical area $\mathrm{A}$ of the Wegner-Wilson loop is given by
$\mathrm{A}=d_\perp\,|z{}^-|\,a_\perp\,a_-$.
Factorization is also true for expectation values of the product of two Wegner-Wilson 
loops which do not overlap. Single plaquette expectation values with respect to the 
ground state wave functional are given by 
\bea
f_{1k}&\equiv& \bra{\Psi_0} \onehalf \Tr{U_{-k}}\ket{\Psi_0}=
\frac{I_2(4\,\rho_0)}{I_1(4\,\rho_0)}+\mathcal{O}(\delta_0^2)\in [-1,1] \nonumber \\
f_{2k}&\equiv& \bra{\Psi_0} \left( \onehalf \Tr{U_{-k}}\right)^2\ket{\Psi_0}=
\frac{I_2(4\,\rho_0)}{4\,\rho_0 \,I_1(4\,\rho_0)}+
\frac{I_3(4\,\rho_0)}{I_1(4\,\rho_0)}+\mathcal{O}(\delta_0^2)\in[0,1] \, ,
\label{eqn:StrongCouplingApproxII}
\eea 
taking into account that higher powers of the same plaquette 
{\it do not factorize}.
Here, $I_n$ denote the modified Bessel functions of the first kind.
The very same relations hold for the purely transversal plaquette expectation 
values with $\rho_0$ substituted by $\delta_0$. By using \eqnref{eqn:StrongCouplingApprox} 
and \eqnref{eqn:StrongCouplingApproxII}, one can analytically evaluate all the gluonic matrix 
elements we need for our calculation. 
One can estimate the physical value of the transversal lattice spacing by identifying the 
rate of the exponential fall-off of a purely transversal Wegner-Wilson loop with the 
dimensionless string tension.
For example, one obtains a transversal lattice spacing of 
$a_\perp\approx 0.65\,\mathrm{fm}$ at $\lambda=10$. This corresponds to a momentum scale 
of $Q^2\approx 1\,\mathrm{GeV}^2$ which is the typical input scale for phenomenological
parameterizations of parton distribution functions.

\section{Modeling a color dipole}
\label{sec:DipoleModel}

The near-light-cone Hamiltonian in \eqnref{eqn:EffectiveLatHamiltonian} contains 
only gluon fields, therefore we cannot {\it derive} hadronic wave functions from this 
Hamiltonian. We have to make a model for the hadron taking into account the gluon 
structure as exactly as possible while treating the quarks only schematically. 
Our model consists of a dipole state with a fixed longitudinal center of mass
momentum $p_-$ localized in transversal configuration space at a fixed 
center of mass position $\vec{b}_\perp=\vec{0}$ while the quark and antiquark 
positions are fixed at $\pm\vec{d}_\perp/2$, i.e. they are separated by
the vector $\vec{d}_\perp$ and connected by a Schwinger string along some
path $\mathcal{C}_\perp$ in the transversal plane specifying the dipole in 
transversal configuration space and longitudinal momentum space
\be
\ket{d(p_-\,;\,-\vec{d}_\perp/2,\mathcal{C}_\perp,\vec{d}_\perp/2)} \; .
\ee
For simplicity, we consider scalar QCD with a scalar matter field. The 
scalar quark fields can be expanded in terms of 
creation and annihilation operators 
\bea
\Phi(x)&=&\Phi_+(x)+\Phi_-(x) \, ,  \\
\Phi_+(x)=\int d\tilde{k}\,a(k)\,
e^{-\mathrm{i}k \, x}~~,~~\Phi_-(x)&=&\int d\tilde{k}\,b^\dagger(k)\,
e^{+\mathrm{i}k \, x}~~,~~d\tilde{k}=\frac{d^4k}{(2\pi)^4}\,2\pi\delta^{(4)}(k^2-m^2) \,. \nonumber
\eea
Here, $\Phi_+(x)$ denotes the positive frequency part of the scalar field and 
$\Phi_-(x)$ represents the negative frequency part. 
The operators $a(k),a^\dagger(k)$ refer to the quark
annihilation/creation operators 
whereas $b(k),b^\dagger(k)$ refer to the antiquark annihilation/creation operators.
$m^2$ denotes the quark/antiquark mass squared. 

In order to construct such a dipole state, 
we start with a dipole state which is localized also in longitudinal configuration space.
Then the dipole state consists of a quark at longitudinal position $x^-$ and at 
transversal position $-\vec{d}_\perp/2$ and of an antiquark at the same $x^-$ and 
at transversal position $\vec{d}_\perp/2$ connected by a Schwinger string along the 
path $\mathcal{C}_\perp$ in the transversal plane in order to achieve gauge invariance.
The transversal path $\mathcal{C}_\perp$ of $n$ steps is parameterized by the intermediate 
transversal positions $\vec{y}_\perp[j]$ ($j=1,n-1$) of the links passed by the path
\be
\mathcal{C}_\perp:-\vec{d}_\perp/2 \to \vec{y}_\perp[1] \to \vec{y}_\perp[2] \to \ldots
\to \vec{y}_\perp[n-1] \to \vec{d}_\perp/2\,.
\ee
The entire dipole state localized in full configuration space is created by some operator
$\chi^\dagger$ acting on the vacuum state 
\be
\ket{\Omega}=\ket{\Phi_0}\otimes\ket{\Psi_0}
\ee
of the entire Hilbert space including gauge fields and (scalar) quarks. 
The vacuum state $\ket{\Phi_0}$ of the (heavy) quark sector is assumed to be the 
Fock vacuum. The vacuum state of the gauge fields, however, is given by \eqnref{eqn:VariationlGSWF}
and therefore of non-perturbative nature. The operator $\chi^\dagger$ has the form
\be
\chi^\dagger 
\,\ket{\Omega}
\Phi_{+}^\dagger(x^-,-\vec{d}_{\perp}/2)\,S_{q\,\bar{q}}^{\mathcal{C}_\perp}(x^-,-\vec{d}_\perp/2\,;\,x^-,\vec{d}_\perp/2)[\{y_j^-\}]\,
\Phi_{-\,}(x^-,\vec{d}_{\perp}/2)\,
\ket{\Omega} \, ,
\label{eqn:DefChiOper}
\ee
where the parallel transporter $S_{q\,\bar{q}}^{\mathcal{C}_\perp}$ 
represents the path ordered ($\mathcal{P}$-ordered) product of $n$ 
transversal link operators along the path $\mathcal{C}_\perp$ in the 
transversal plane. 

\begin{figure}
\centering
\includegraphics[width=0.8\textwidth]{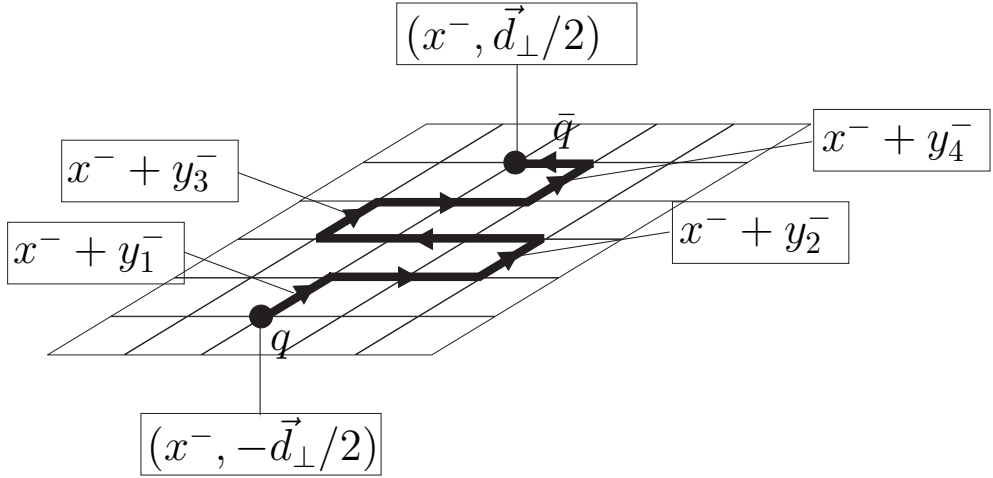}
\caption{\label{fig:Dipole} Graphical representation of the dipole
  state on the lattice. The black dots represent the quark and the antiquark.
  The transversal links are allowed to move freely along the 
  longitudinal direction. For simplification only transversal links in one direction
  are shown. 
  The coordinates $y_j^-$ denote the displacement of the link
  $j$ along the longitudinal direction with respect to the longitudinal position of the
  quark and antiquark.}
\end{figure}

In addition, we have allowed in \eqnref{eqn:DefChiOper} 
for different longitudinal positions of the 
transversal links as motivated below and have inserted
longitudinal Schwinger string bits connecting the 
(otherwise adjacent) transversal links
in order to retain gauge invariance
\bea
\lefteqn{S_{q\,\bar{q}}^{\mathrm{C}_\perp}(x^-,-\vec{d}_\perp/2\,;\,x^-,\vec{d}_\perp/2)[\{y_j^-\}]
\equiv
S_-(x^-,-\vec{d}_\perp/2;y_1^-,-\vec{d}_\perp/2)
\,S_\perp(y_1^-,-\vec{d}_\perp/2\,;\,y_1^-,\vec{y}_\perp[1])}  
\nonumber \\ 
& &
\,\left[\mathcal{P}\prod\limits_{j=1}^{n-1} 
S_-(y_j^-,\vec{y}_\perp[j];\,y_{j+1}^-,\vec{y}_\perp[j])
S_\perp(y_{j+1}^-,\vec{y}_\perp[j]\,;\,y_{j+1}^-,\vec{y}_\perp[j+1])\right]
\, 
S_-(y_n^-,\vec{d}_\perp/2\,;\,x^-,\vec{d}_\perp/2) \, . \nonumber \\
& &\label{eqn:DefConn}
\eea
Here, $\vec{y}_\perp[n]=\vec{d}_\perp/2$.
The argument in round brackets of $S_{q\,\bar{q}}^{\mathcal{C}_\perp}$ represents the 
starting and the end point of the wiggly string,
whereas the argument in square brackets represents the set of longitudinal coordinates
of the intermediate transversal links. 
A typical string configuration is graphically represented in
\figref{fig:Dipole} assuming a lattice structure. Each 
of the link operators represents a string bit of the entire 
string.
The transversal part of the transporter 
between the quark and the antiquark along the 
vector $\vec{d}$ with minimal length (in purely transversal direction) 
represents the ground 
state of the nlc Hamiltonian in the strong coupling limit. 
In this limit, the nlc Hamiltonian is dominated by the chromo-electric field
operators and the energy of the dipole state scales with the transversal 
length $|{\vec d}|$ of the gluonic string. 
Because of the Lorentz boost in the longitudinal
direction accompanied by the transition from lab frame coordinates to nlc 
coordinates, the transverse electric field strength operators appear in the 
near-light-cone Hamiltonian with a weight larger by a factor 
$1/\tilde{\eta}^2$ 
compared to the longitudinal field strengths (c.f. \eqnref{eqn:EffectiveLatHamiltonian}).
Therefore every string with a fixed number of transversal links practically
has the same energy regardless of the number of links in $x^-$-direction in the light 
cone limit $\tilde{\eta} \to 0$. This implies that the string 
should be defined having any number of longitudinal links. 

The aim of our calculation is to calculate the gluon structure function for a 
color dipole with string configurations deformed in this manner, however with fixed momentum. 
We have to construct from the localized dipole in configuration space a fast moving momentum 
eigenstate with total longitudinal momentum $p_-$ for which we can determine the momenta of 
gluons. For this purpose we perform an integration over all translations of this state over 
the entire coordinate range multiplied with the appropriate momentum eigenfunctions 
$e^{-\mathrm{i}\,p_-\,x^-}$.
Finally, weighting will be performed with the vacuum wave functional squared.
Since none of the link configurations is preferable from the energetic point of 
view, we integrate over all possible link configurations and assign to each configuration 
a probability amplitude $\Psi(\{y_j^-\})$. We model $\Psi(\{y_j^-\})$ by the product
of momentum eigenfunctions of each link integrated over all possible link 
momenta. Due to the projection of the entire dipole state onto total momentum $p_-$,
the sum of its constituent momenta is restricted (c.f. \eqnref{eqn:posLatMom}), i.e. 
\be
\Psi(\{y_j^-\})=
\int\limits_{0}^{\infty} \left(\prod\limits_{j=1}^n dl_-^{j}\right) \,e^{-\mathrm{i}\,
\sum\limits_{j=1}^n l_-^{j}\,y_j^-}\,\Theta\left(p_--\sum_{j=1}^n l_-^j\right) \; .
\ee
Hence, the final dipole state is given by
\bea 
\lefteqn{
\ket{d(p_-\,;\,-\vec{d}_{\bot}/2,\mathrm{C}_\perp,\vec{d}_\perp/2)}=
\frac{1}{\sqrt{N}}\int \left(\prod\limits_{j=1}^n
dl_-^{j}dy_j^-\right)\,e^{-\mathrm{i}\,
\sum\limits_{j=1}^n l_-^{j}\,y_j^-}\,\Theta\left(p_--\sum_{j=1}^n l_-^j\right)}
\label{eqn:DWLP}
\label{eqn:FinalDipoleState}
\\
& &\times \int dx^-\,e^{-\mathrm{i}\,p_-\,x^-}\,
\Phi_{+}^\dagger(x^-,-\vec{d}_{\perp}/2)\,
S_{q\,\bar{q}}^{\mathrm{C}_\perp}(x^-,-\vec{d}_\perp/2\,;\,x^-,
\vec{d}_\perp/2)[\{x^-+y_j^-\}]\, 
\Phi_{-}(x^-,\vec{d}_{\perp}/2)\,
\ket{\Omega}\,.\nn
\eea
The appropriate normalization of the dipole state is guaranteed by division with a 
suitable normalization factor $\sqrt{N}$. This dipole state represents the starting 
point for our investigation of its gluon structure. 

\begin{figure}
\centering
\includegraphics[width=.8\textwidth]{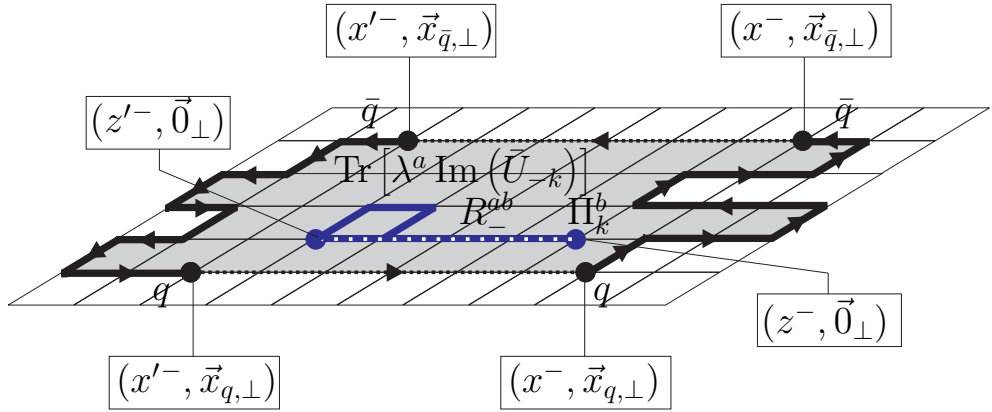}
\caption{\label{fig:EVOPD} Graphical representation of the generalized Wegner-Wilson
  loop generated by the two color dipole states with a quark $q$ and an
  antiquark $\bar q$ connected by n-links in transversal direction. For simplification
  only transversal links in one direction are shown.  
  The dash-dotted insertion in the so-formed Wegner-Wilson loop represents the
  gluonic two-point operator
   with a lattice electric field operator at longitudinal
  coordinate $z^-$ and a lattice magnetic 
  field operator at longitudinal coordinate $z^\prime{}^-$.}
\end{figure}

Matrix elements between two dipole states can be computed by 
contracting the scalar operators $\Phi_{\pm}$ yielding the Feynman propagator $\Delta_F(x,y|A)$.
We find for the Feynman propagator of the interacting
scalar theory in the eikonal approximation (quark/antiquark have large $p_+$ momenta) 
\bea
\Delta_F(x,y|A)&=&\frac{p_-}{m^2}\,\left[\Theta(x^--y^-)\,
e^{-\mathrm{i}\,p_-\,(x^--y^-)}\,+\Theta(y^--x^-)\,e^{-\mathrm{i}\,p_-\,(y^--x^-)}\right] \nonumber \\
& &\times S_-(x^-,\vec{x}_\perp ~;~y^-,\vec{x}_\perp)\,\delta^{(2)}(\vec{x}_\perp-\vec{y}_\perp)\,\delta(x^+-y^+)\,.
\label{eqn:PropMT}
\eea
At high $p_+$-momentum, the quark and antiquark in the color dipole move on
straight line classical trajectories and pick up non-abelian phase factors along
their paths. Instead of the usual time ordering, we have an ordering along the longitudinal 
spatial coordinate. Thus, by evaluating matrix elements between two dipole states,  
additional straight line Schwinger strings appear along the longitudinal 
direction connecting the wiggly Schwinger strings from the incoming and outgoing dipole states.
To evaluate the expectation value of the point split operator $G$ 
(given in \eqnref{eqn:nlcCorrFunc}) between two dipole states with fixed string configurations 
the following expression can be reduced to a purely gluonic matrix element
\bea
& &
\bra{\Omega}
\Big(\Phi_{+}^\dagger(x^\prime{}^-,\vec{x}_\perp{}^\prime)\,
S_{q\bar{q}}^{\mathrm{C}_\perp^\prime}\big(x^\prime{}^-,\,\vec{x}_\perp{}^\prime\,;\,x^\prime{}^-,\,\vec{x}_\perp{}^\prime+\vec{d}_\perp{}^\prime\big)[\{y_j^\prime{}^-\}]\,
\Phi_{-}(x^\prime{}^-,\vec{x}_\perp{}^\prime+\vec{d}_\perp{}^\prime)\Big)^\dagger 
 \nonumber \\
& &~~~~~~~~\times G\,\Big(\Phi_{+}^\dagger(x^-,\vec{x}_\perp)\,
S_{q\bar{q}}^{\mathrm{C}_\perp}\big(x^-,\vec{x}_\perp\,;\,x^-,\vec{x}_\perp+\vec{d}_\perp\big)[\{y_j^-\}]\,
\Phi_{-}(x^-,\vec{x}_\perp+\vec{d}_\perp)\Big) \ket{\Omega} \nonumber \\
&=&\bra{\Psi_0} \mathrm{Tr}\Big[ S_-(x^-,\vec{x}_\perp+\vec{d}_\perp\,;\,x^{\prime}{}^-;\vec{x}_\perp+\vec{d}_\perp)\,
S_{q\bar{q}}^{\mathrm{C}_\perp^\prime}\big(x^\prime{}^-,\vec{x}_\perp\,;x^\prime{}^-,\vec{x}_\perp+\vec{d}_\perp\,\big)[\{y_j^\prime{}^-\}]^\dagger \nonumber \\
& & ~~~~~~~~~~~~ \times
S_-(x^\prime{}^-,\vec{x}_\perp\,;\,x^-;\vec{x}_\perp)\,
G \, S_{q\bar{q}}^{\mathrm{C}_\perp}\big(x^-,\vec{x}_\perp\,;\,x^-,\vec{x}_\perp+\vec{d}_\perp\big)[\{y_j^-\}]  \Big]\ket{\Psi_0} \nonumber \\
& & ~~~~~\times \frac{p_-^q\,p_-^{\bar{q}}}{m^4}\,e^{\mathrm{i}(p_-^q+p_-^{\bar{q}})(x^--x^\prime{}^-)}\, \delta^{(2)}(\vec{x}_\perp-\vec{x}_\perp{}^\prime)\,\delta^{(2)}(\vec{d}_\perp-\vec{d}_\perp{}^\prime)   
\label{eqn:IntegratedFermions}
\eea
to be obtained by averaging over the vacuum wave functional squared.
The transversal chromo-electric field operators appearing in \eqnref{eqn:nlcCorrFunc} 
do not commute with the transversal link operators appearing in the definition of the 
dipole operator. Therefore, one has to take care of the right arrangement of the operators.  
The string $S_{q\bar{q}}^{\mathcal{C}_\perp^\prime\,\dagger}(x^{\prime\,-})$ arising from
the dipole at $x^\prime{}^-$ must appear to the left of the
operator $G(z^-,\vec{z}_\perp\,;\,z^-{}^\prime,\vec{z}_\perp)$ and correspondingly the string
$S_{q\bar{q}}^{\mathcal{C}_\perp}(x^-)$ 
to the right of $G(z^-,\vec{z}_\perp\,;\,z^-{}^\prime,\vec{z}_\perp)$ in the matrix element.
The resulting \eqnref{eqn:IntegratedFermions} allows to express the expectation value of 
the momentum density operator between dipole states by 
the gluonic vacuum average of the trace over a non-rectangular 
Wegner-Wilson loop whose edges are given by the transversal parallel transporters 
$S_{q\bar{q}}^{\mathcal{C}_\perp}$, $S_{q\bar{q}}^{\mathcal{C}_\perp^\prime\,\dagger}$ 
and
the longitudinal straight line Schwinger strings connecting the two dipole
states. Such a Wilson loop is shown in \figref{fig:EVOPD}, for simplification
with only one transverse dimension and the $x^-$-direction. 
The full curves represent the strings which connect the quark and antiquark in each of 
the dipole states. The dotted strings arise due to the elimination of
the quark/antiquark operators in the eikonal approximation of the quark propagator.
The blue curve corresponds to the point split operator $G$. Note, that in general the 
wiggly string can extend into both transverse directions and the $x^-$-direction.
After the quark fields are eliminated, the $x^-$ integration in \eqnref{eqn:FinalDipoleState} 
can be performed for the incoming and outgoing dipole state and yields together with the quark 
and antiquark momenta from \eqnref{eqn:IntegratedFermions} a delta distribution setting the sum of transversal
link momenta equal to the total string momentum $p_-^S$  
\be
\delta\left(p_-^S-\sum_{j=1}^n l_-^j\right)~~,~~
p_-^S=p_--p_-^q-p_-^{\bar{q}}\,.
\ee
In order to compute the normalization of the dipole state, one simply has to substitute
the point split operator $G$ in \eqnref{eqn:IntegratedFermions} by the unit operator.

\section{Near-light-cone gluon correlation function on the \\ lattice}
\label{sec:NLCCorrLat}

We come now to the practical evaluation of the lattice counterpart of the gluon
distribution function. 
On the lattice, a direct simulation of the average gluon momentum fraction becomes subtle.
Being the lattice generator of longitudinal translations, the
longitudinal momentum operator induces translations by any multiple of the lattice unit. 
We discriminate between longitudinal lattice momenta and eigenvalues of the longitudinal 
lattice momentum operator in the following.
Similar to the 
dispersion relation for fermions on the lattice, 
one has for each positive valued eigenvalue in the spectrum of the longitudinal lattice
momentum operator two possible lattice momenta corresponding to this eigenvalue. Even worse, the largest
possible lattice momentum corresponds to an eigenvalue of the longitudinal momentum operator close
to zero, far away from its maximal possible value. 
Hence, by choosing the hadron to have the maximal lattice momentum, one finds gluon momentum fractions
which do not add up to unity (neglecting quark momenta).

We discuss first how the problem arises and second how to circumvent it. 
We start discretizing the point split operator of the  
nlc correlation function \eqnref{eqn:nlcCorrFunc} :
\bea
\lefteqn{
G(z^-,\vec{z}_\perp~;~z^\prime{}^-,\vec{z}_\perp)} \nonumber \\
&=&\frac{1}{4}\,\sum\limits_k \Big(2\,\Pi_k^a(z^-,\vec{z}_\perp)\,
S_{a b}^A(z^-,\vec{z}_\perp~;~z^\prime{}^-,\vec{z}_\perp)\,
\Tr{\frac{\sigma^b}{2}\,\Imag{\overline{U}_{-k}(z^\prime{}^-,\vec{z}_\perp)}}
\label{eqn:NLCCorrFuncLattFine}\\
& &~~~~~~~~~~~+2\,\Pi_k^a(z^\prime{}^-,\vec{z}_\perp)\,
S_{a b}^A(z^\prime{}^-,\vec{z}_\perp,z^-,\vec{z}_\perp)\,
\Tr{\frac{\sigma^b}{2}\,\Imag{\overline{U}_{-k}(z^-,\vec{z}_\perp)}}+h.c.\Big) \, .\nonumber 
\eea
The field strength $F_{-k}^a$ in lattice form is expressed here
in such a way 
that \eqnref{eqn:nlcCorrFunc} follows in the naive continuum limit. Here,
$\overline{U}_{-k}(\vec{x})$ is an average over the "forward"
plaquette $U_{-k}^{\rm forw}(\vec{x})$ right of $x^-$ and of the "backward" plaquette
$U_{-k}^{\rm backw}(\vec{x})$ left of $x^-$, both $-k$ plaquettes 
beginning in $x$  and adjacent to the transversal link $U_{k}(\vec{x})$ :
\be 
\overline{U}_{-k}(\vec{x})\equiv
\onehalf\left( U_{-k}^{\rm forw}(\vec{x})+U_{-k}^{\rm backw}(\vec{x}) \right)
\ee
with
\be
U_{-k}^{\rm forw}(\vec{x})\equiv
U_{-}(\vec{x}) U_{k}(\vec{x}+\wh{e}_-)U_{-}^{\dagger}(\vec{x}+\wh{e}_k)U_{k}^{\dagger}(\vec{x})
\ee
and
\be
U_{-k}^{\rm backw}(\vec{x})\equiv
U_{k}(\vec{x})U_{-}^\dagger(\vec{x}+\wh{e}_k-\wh{e}_-)
U_{k}^{\dagger}(\vec{x}-\wh{e}_-)U_{-}(\vec{x}-\wh{e}_-) \; .
\ee
Note, that the orientations of the forward and backward plaquettes are the
same such that the projection of the traceless antihermitean part $\mathrm{Im}\,\overline{U}_{-k}$ 
onto $\sigma^a/2$ becomes proportional to $F_{-k}^a(\vec{x})$ in the continuum limit.

For $z^\prime{}^-=z^-$, the point split operator 
reduces to the dimensionless nlc momentum density operator 
(c.f. \eqnref{eqn:DefLongMomDens}) with $\vec{z}=(z^-,\vec{z}_\perp)$, the lattice form of which is  
\bea
\mathcal{P}_-(\vec{z})&=&
\sum\limits_k\onehalf
\Big(2\,\Pi_k^a(\vec{z})\,\Tr{\frac{\sigma^a}{2}\,\Imag{\overline{U}_{-k}(\vec{z})}} \nonumber \\
& & ~~~~~~~~~~~~~~
+2\,\Tr{\frac{\sigma^a}{2}\,\Imag{\overline{U}_{-k}(\vec{z})}}\Pi_k^a(\vec{z})
\Big) \, ,
\label{eqn:LMomDensLat}
\eea 
which becomes the total longitudinal momentum operator $P_-$ when
summed over the entire lattice. The variational ground state wave functional 
ansatz \eqnref{eqn:VariationlGSWF} is an exact eigen state of the longitudinal momentum operator with eigen value
equal to zero for $\delta_0=0$, i.e.
\be
\sum_{\vec{z}} \mathcal{P}_-(\vec{z})\,\ket{\Psi_0}=0+\mathcal{O}(\delta_0) \, .
\ee
To find the spectrum of the total longitudinal momentum operator, one has
to know how it acts on link operators.
The commutator of the total momentum operator 
$P_- = \sum_{\vec{z}} \mathcal{P}_-(\vec{z})$ with a transversal 
link $U_j(\vec{y})$ 
being part of the gluonic string forming the hadron state 
gives (c.f. \eqnref{eqn:cancommrellat}) 
\bea
\left[\sum_{\vec{z}} \mathcal{P}_-(\vec{z})
,U_j(\vec{y})\right]&=&\frac{1}{4\,\mathrm{i}}\left(
U_-(\vec{y})\,U_j(\vec{y}+\vec{e}_-)\,U^\dagger_-(\vec{y}+\vec{e}_j)-
U_{-j}^{{\rm forw}\,\dagger}(\vec{y})\,U_j(\vec{y})+U_{-j}^{\rm backw}(\vec{y})\,U_j(\vec{y})\right.\nonumber \\
& & ~~~~~~~~~~\left.-
U_-^\dagger(\vec{y}-\vec{e}_-)\,U_j(\vec{y}-\vec{e}_-)\,U_-(\vec{y}+\vec{e}_j-\vec{e}_-)
\right) \nonumber \\ &=&
\frac{1}{2\,\mathrm{i}}\left(U_-(\vec{y})\,U_j(\vec{y}+\vec{e}_-)\,U^\dagger_-(\vec{y}+\vec{e}_j)\right. \nonumber \\
& &\left.~~~~~-
U_-^\dagger(\vec{y}-\vec{e}_-)\,U_j(\vec{y}-\vec{e}_-)\,U_-(\vec{y}+\vec{e}_j-\vec{e}_-)
\right)+\mathcal{O}(a^2) \, .
\label{eqn:CommPmLink}
\eea
The first line of \eqnref{eqn:CommPmLink} 
(containing also ``curled-up'' plaquette
insertions into the gluonic string) is exact and will be used in subsequent 
calculations. The exact commutator of $P_-$ with the link $U_j$ symbolized
by the arrow $\link$ has the following graphical representation
\bea
\left[P_-,\link \right]=\frac{1}{4\,\mathrm{i}}\left(\stapler-\staplefp+\staplebp-\staplel\right)\,.
\eea
In the second line of \eqnref{eqn:CommPmLink}, we have expanded the result in powers of 
the lattice spacing up to quadratic corrections $\mathcal{O}(a^2)$, where $a^2$
stands for $(a_-^2,a_-\,a_\perp,a_\perp^2)$. 
Then the commutator of $P_-$ with the link $U_j$ corresponds 
to gauge invariant forward and backward translations 
along the longitudinal direction  
\bea
\left[P_-,\link \right]=\frac{1}{2\,\mathrm{i}}\left(\stapler-\staplel\right)+\mathcal{O}(a^2)\,,
\eea
i.e. leads to a discretized covariant 
first derivative of the link implemented in a symmetric way. 
The Heisenberg equation of motion for the transversal link 
on the lattice identifies the longitudinal momentum operator as the generator of longitudinal 
translations. In $A_-=0$ gauge, i.e. for $U_-=\unitop$ the covariant derivative reduces to an ordinary discretized derivative.

Eigenstates of the longitudinal momentum operator can be found as sums of transverse links along the
longitudinal direction modulated by appropriate phase factors up to corrections quadratic in the lattice
constants $(a_-,a_\perp)$ 
\be
\left[\sum_{\vec{z}} \mathcal{P}_-(\vec{z})
,V_j(p_-,\vec{y}_\perp)\right]=\sin(p_-)\,V_j(p_-,\vec{y}_\perp)+\mathcal{O}(a^2)
\label{eqn:MomEigenVal}
\ee
with
\be
V_j(p_-,\vec{y}_\perp)=\sum_{y^-}e^{-\mathrm{i}\,p_-\,y^-}\,U_j(y^-,\vec{y}_\perp) \, .
\ee
They arise from projecting transversal links localized in configuration space onto
a definite longitudinal momentum $p_-$ and are
not elements of $SU(2)$ because they are superpositions
of link operators. 
The longitudinal lattice momenta must be
an integer multiple $n$ of $2\,\pi/N_-$ 
with $n\leq 0 \leq N_-/2-1$, since
the longitudinal light cone momentum for an on shell particle 
is always positive (c.f. \eqnref{eqn:posLatMom}). 
The momentum $p_-$ of the target is chosen as
the largest momentum in order to have the maximum resolution in the gluon distribution
function \cite{Pauli:1985ps,Burkardt:2001jg}
\be 
p_-=\frac{2\,\pi}{N_-}(N_-/2-1) \, .  
\label{eqn:LargestLatticeMom}
\ee
Longitudinal lattice gluon momenta have the resolution
\be 
\frac{\Delta p_-^g}{p_-}=\frac{2}{N_--2}\,.  
\ee 
In order to have a
high resolution, the extension of the lattice in the longitudinal
direction has to be very large. 
 
Eqs.~(\ref{eqn:MomEigenVal},\ref{eqn:LargestLatticeMom}) imply that the largest lattice momentum $p_-$ yields an eigenvalue
of the longitudinal momentum operator approximately equal to zero.
Even though the gauge field is a bosonic degree of freedom, the eigen value $\sin(p_-)$ 
of the discretized momentum operator looks ''fermionic'' in the Brillouin zone $p_-\in[-\pi,\pi]$ 
of the longitudinal momentum, i.e. the map $p_- \to \sin(p_-)$ is not injective.

In order to make it injective and monotonically increasing, we perform a similar but much simpler operation as  
for Kogut Susskind fermions, i.e. we block links 
on a sublattice with half of the lattice spacing along the longitudinal direction. 
Even sites on the fine lattice can be identified with lattice sites on the 
original lattice. Odd lattice sites on the fine lattice lie between two 
neighbouring original lattice sites. 
The physical extension $L_{\mathrm{phys}}$ and the physical momenta are kept fixed during the transition from the original to the fine lattice along the longitudinal direction 
\bea
L_{\mathrm{phys}} &=&N_-\,a_- \, , \nonumber \\
N_-&=&\frac{N_-^f}{2} \, , \nonumber \\
a_-&=&2\,a_-^f \, , \nonumber \\
p_-^f&=&\frac{p_-}{2} \, .
\label{eqn:ConvOrigFineLat}
\eea
We denote quantities on the fine sublattice by 
superscripts $f$.
Using the fine lattice, 
we define a new momentum eigenstate $\widetilde{V}_j(p_-,\vec{y}_\perp)$ 
on the original lattice by a modulated sum over fine lattice links $U_j^f(y_f^-,\vec{y}_\perp)$
\be
\widetilde{V}_j(p_-,\vec{y}_\perp)
=\sum_{y^-_f}e^{-\mathrm{i}\,p_-^f\,y^-_f} \, \left.
U_j^f(y_f^-,\vec{y}_\perp)\right|_{p_-^f=p_-/2} \, .   
\label{eqn:MESCF}                          
\ee
Thus, by keeping the physical momenta fixed, the allowed fine lattice momenta are 
given by one half of the original lattice momenta. Thereby we reduce the possible lattice momenta
on the fine lattice by a factor of two and obtain a one-to-one correspondence between the 
lattice momenta and the eigenvalues on the fine lattice.
The right hand side of \eqnref{eqn:MESCF} is obviously an eigenstate of the 
longitudinal momentum operator on the fine lattice albeit with eigenvalue $\sin(p_-^f)$.
However, \eqnref{eqn:MESCF} is not an eigenstate of the momentum operator
on the original lattice, 
because the original momentum operator applied to the block 
averaged state does act solely 
on fine lattice links at even longitudinal fine lattice sites.
Therefore, we have to introduce a block averaged 
longitudinal momentum density $\widetilde{\mathcal{P}}_-$ on the original lattice,
acting on even and odd fine lattice sites.
It is given by the following sum of fine momentum density operators 
$\mathcal{P}_-^f(z^-_f,\vec{z}_\perp)$ which are defined as in \eqnref{eqn:LMomDensLat} with all operators on the fine lattice ($\Pi_k^a\rightarrow\Pi_k^{a\,f}$ and $U_j\rightarrow U_j^f$)
\be 
\widetilde{\mathcal{P}}_-(z^-,\vec{z}_\perp)=2\,
\left(\onehalf\mathcal{P}_-^f(2\,z^--1,\vec{z}_\perp)+
\mathcal{P}_-^f(2\,z^-,\vec{z}_\perp)+
\onehalf\mathcal{P}_-^f(2\,z^-+1,\vec{z}_\perp)\right) \, .
\label{eqn:LongMomDensCoarse}
\ee
The factor two in front of the definition
originates from converting the fine lattice  operator
into an operator on the coarse lattice, i.e. $P_-=2\,P_-^f$ similar to \eqnref{eqn:ConvOrigFineLat}. 
The effective longitudinal momentum
density on the original lattice \eqnref{eqn:LongMomDensCoarse} has  
contributions from even and odd sites on the fine lattice. Since we have 
symmetrized the operator with respect to the odd lattice sites on the fine lattice by
using one half of the forward and one half of the backward
contribution, $\widetilde{V}_j(p_-,\vec{y}_\perp)$ is an
eigenstate of the effective longitudinal momentum operator on the original lattice
with eigenvalue 
\be
\left[\widetilde{P}_-,\widetilde{V}_j(p_-,\vec{y}_\perp)\right]
=2\,\sin(p_-/2)\,\widetilde{V}_j(p_-,\vec{y}_\perp)+\mathcal{O}(a_f^2) \, .
\label{eqn:LongMomC}
\ee
Now, the largest possible lattice momentum does also correspond to the largest possible
eigenvalue of the momentum operator and the eigenvalues are monotonically increasing.
The above expression \eqnref{eqn:LongMomC} of the longitudinal momentum also appears in the dispersion
relation for bosons $\omega=\sqrt{\sum_i (2\,\sin(p_i/2))^2+M^2}$ and defines a 
one-to-one mapping of the lattice momenta $p_i$ in the first Brillouin zone to the
energy states. 
It represents an important 
stratification which allows to calculate momentum
fractions.  
We define an effective correlation function $\widetilde{G}(z^-,\vec{z}_\perp\,;\,z^-{}^\prime,\vec{z}_\perp)$ 
on the original lattice by averaging the correlation functions $G^f(z^-_f,\vec{z}_\perp\,;\,z^-_f{}^\prime,\vec{z}_\perp)$
on the fine lattice \eqnref{eqn:NLCCorrFuncLattFine}
\bea
\widetilde{G}(z^-,\vec{z}_\perp\,;\,z^-{}^\prime,\vec{z}_\perp)
&=&2\,\left(\onehalf G^f(2\,z^--1,\vec{z}_\perp\,;\,2\,z^-{}^\prime-1,\vec{z}_\perp)
+G^f(2\,z^-,\vec{z}_\perp\,;\,2\,z^-{}^\prime,\vec{z}_\perp)\right. \nonumber \\
& &\left.
+\onehalf G^f(2\,z^-+1,\vec{z}_\perp\,;\,2\,z^-{}^\prime+1,\vec{z}_\perp)\right) \, .
\label{eqn:NLCCorrFuncLattCoarse}
\eea
This definition is in agreement with the longitudinal
momentum density operator \eqnref{eqn:LongMomDensCoarse} on the coarse lattice for $z^-=z^-{}^\prime$.
Finally, the lattice definition of the gluon distribution function is given by
\bea
\lefteqn{
g(p_-^{g})=\lim_{\eta\rightarrow 0}\, \frac{1}{\pi}\,\frac{1}{p_-^{g}}
\sum_{z^-,z^\prime{}^-}\,\sum_{\vec{z}_\perp}
e^{-\mathrm{i}\,p_-^{g}(z^--z^\prime{}^-)}}\label{eqn:EffLatGluonStruc}\\
& & 
\frac{
\bra{d(p_-\,;\,-\vec{d}_\perp,\mathcal{C}_\perp,\vec{d}_\perp/2) }\widetilde{G}(z^-,\vec{z}_\perp~;~z^\prime{}^-,\vec{z}_\perp)
\ket{d(p_-\,;\,-\vec{d}_\perp,\mathcal{C}_\perp,\vec{d}_\perp/2)}_c}{\left\langle d(p_-\,;\,-\vec{d}_\perp,\mathcal{C}_\perp,\vec{d}_\perp/2) \right. \left| d(p_-\,;\,-\vec{d}_\perp,\mathcal{C}_\perp,\vec{d}_\perp/2) \right\rangle}\,.\nn  
\label{eqn:GluonDistLatCoarse}
\eea
In order not automatically to enforce Bjorken scaling,
we prefer to express the gluon distribution function in terms of the gluon 
momentum $p_-^{g}$ instead of the momentum fraction $x_B$.
Here, the gluonic component of the hadronic target state has to be defined by the 
block averaged momentum eigenstates given in \eqnref{eqn:MESCF}.

On the lattice, the following orthogonality relation holds for positive gluon momenta
\be
\frac{1}{N_-}\sum_{p_-^{g}=0}^{p_-} \mathrm{Re}\left[e^{-\mathrm{i}\,p_-^g\,z^-}\right]=\onehalf \delta_{z^-,0}+\frac{1}{2\,N_-}\left(1-(-1)^{z^-} \right)=\onehalf\delta_{z^-,0}+\mathcal{O}\left(\frac{1}{N_-} \right) \, .
\ee
Taking the real part in the orthogonality relation is sufficient since the point split 
operator is symmetric with respect to an interchange of $z^-$ and $z^\prime{}^-$.
In addition to the continuum result $\delta_{z^-,0}/2$, there is also a finite size contribution which vanishes like $1/N_-$. 
One finds the average gluon momentum $\langle p_-^{g} \rangle$ 
\bea
\left\langle p_-^{g} \right\rangle 
&=&\frac{2\,\pi}{N_-}\sum_{p_-^{g}=0}^{p_-}p_-^{g}\,g(p_-^{g})=\sum_{z^-,\vec{z}_\perp} \left\langle \widetilde{\mathcal{P}}_- \right\rangle +\mathcal{O}
\left( \frac{1}{N_-} \right) \, , \nonumber \\
\sum_{z^-,\vec{z}_\perp}\left\langle  \widetilde{\mathcal{P}}_- \right\rangle
&=&\sum_{z^-}\sum_{\vec{z}_\perp}
\frac{\bra{d(p_-\,;\,-\vec{d}_\perp,\mathcal{C}_\perp,\vec{d}_\perp/2) }\widetilde{\mathcal{P}}_{-}(z^-,\vec{z}_\perp)
\ket{d(p_-\,;\,-\vec{d}_\perp,\mathcal{C}_\perp,\vec{d}_\perp/2)}_c}{\left\langle d(p_-\,;\,-\vec{d}_\perp,\mathcal{C}_\perp,\vec{d}_\perp/2) \right. \left| d(p_-\,;\,-\vec{d}_\perp,\mathcal{C}_\perp,\vec{d}_\perp/2) \right\rangle} \,
\label{eqn:MomSumRuleLat} \\
&=& \sum_{z^-}\sum_{\vec{z}_\perp} \frac{
\bra{d(p_-\,;\,-\vec{d}_\perp,\mathcal{C}_\perp,\vec{d}_\perp/2) }\widetilde{G}(z^-,\vec{z}_\perp~;~z^-,\vec{z}_\perp)
\ket{d(p_-\,;\,-\vec{d}_\perp,\mathcal{C}_\perp,\vec{d}_\perp/2)}_c}{\left\langle d(p_-\,;\,-\vec{d}_\perp,\mathcal{C}_\perp,\vec{d}_\perp/2) \right. \left| d(p_-\,;\,-\vec{d}_\perp,\mathcal{C}_\perp,\vec{d}_\perp/2) \right\rangle}\,
. \nonumber
\eea
Here, the factor $2\,\pi/N_-$ is due to the discretized measure of the momentum integration.
In the infinite volume limit, the leading contribution is of order $\mathcal{O}(1)$ due to the
normalization of the dipole state.

\section{Gluon structure function of a one-link dipole}
\label{sec:GluonDistSingleLink}

We start with the computation of the gluon structure function for the one-link dipole. 
Later we will consider the gluon structure 
function of the more sophisticated multi-link dipole.
For a computation we use the
dipole state \eqnref{eqn:FinalDipoleState} 
reduced to a single link. 
Since the pure glue Hamiltonian of \eqnref{eqn:EffectiveLatHamiltonian} does not control quark 
dynamics we have to choose between two alternatives:
\begin{itemize} 
\item let the quark and antiquark simply follow the
gluon link to which they are attached to and fix the quark
and antiquark momenta to the correspondent
link momentum.   
\item impose the quark dynamics
of the color dipole externally. Since the total
hadron longitudinal momentum is given by the sum of the momenta of its constituents, the total gluon momentum is then fixed. 
\end{itemize}
We follow the second alternative
and take the mean gluon momentum from experiment. At the input scale $Q^2\approx\pi^2/a_\perp^2= 1.5\,\mathrm{GeV}^2$ corresponding to $\lambda\approx10$ (c.f. \secref{sec:NLCHamiltonian}), 
we use the MRST-parameterization \cite{Martin:2001es} and assign a mean momentum fraction
$p_-^S=0.38\,p_-$ to the string.
The string momentum $p_-^S$ results from the difference of hadron momentum and 
quark and the antiquark momenta which is taken from experiment:
\be
p_-^S=p_--p_-^q-p_-^{\bar{q}} \, .
\ee
We ascribe this momentum $p_-^S$ to the complete string of gluon links. 
Its transverse size $\vec{d}_{\bot}$
now equals one of the lattice unit vector $\vec{e}_{j}$, where $j=1,2$ denote 
the transversal directions
\be
\vec{d}_{\bot}=\vec{e}_{j}~~,~~j=1,2\,.
\ee 
According to \eqnref{eqn:IntegratedFermions},
the norm of the one-link dipole is related to the matrix element of a Wegner-Wilson loop defined by 
the eikonal trajectories $S_-^f,S_-^f{}^\dagger$ of the quark and antiquark (dotted lines) together with
the strings (full lines) $S_{q\bar{q}}^{f\,\dagger}[x_f^\prime]$, $S_{q\bar{q}}^f[x_f^-]$ inside of 
the color dipole (c.f. \figref{fig:EVOPD}). 
For a string consisting of a single link, we can not
define the dipole state 
symmetrically with respect to the origin in transversal space.
We choose without loss of generality the quark to be located at the origin in transversal space and
extend the dipole along one of the positive transversal axes with the antiquark located
at $\vec{e}_j$. Hence, the norm of the dipole state is given by
\bea
\lefteqn{
\bra{d(p_-\,;\,\vec{0}_{\bot},\vec{d}_\perp)}\left. d(p_-\,;\,\vec{0}_{\bot},\vec{d}_\perp)
\right\rangle =
\frac{1}{N}\sum_{x_f^-,x_f^\prime{}^-} 
e^{-\mathrm{i}\,p_-^S/2\, \left(x_f^--x_f^\prime{}^-\right)}\bra{\Psi_0}\mathrm{Tr}\Big[ S_-^f(x_f^-,\vec{d}_\perp\,;\,x_f^{\prime}{}^-;\vec{d}_\perp)\phantom{S_{q\bar{q}}^{f\,\dagger}\big(x_f^\prime{}^-,\vec{0}_\perp\,\big)}} \nonumber \\
& & \hspace{0.cm}
\times \,
S_{q\bar{q}}^{f\,\dagger}\big(x_f^\prime{}^-,\vec{0}_\perp\,;\,x_f^\prime{}^-,\vec{d}_\perp\big)[x_f^\prime{}^-]\, 
S_-^f(x_f^\prime{}^-,\vec{0}_\perp\,;\,x_f^-,\vec{0}_\perp)\,
S_{q\bar{q}}^f\big(x_f^-,\vec{0}_\perp\,;\,x_f^-,\vec{d}_\perp\big)[x_f^-]\Big]\ket{\Psi_0}\,.
\label{eqn:DefNormDipole}
\eea
Here, we have suppressed the path index $\mathcal{C}_\perp$ in the state vectors.
We absorb factors $2\,\pi$, volume factors due to the 
squared delta distributions and the factor $p_-^q\,p_-^{\bar{q}}/m^4$ appearing in \eqnref{eqn:IntegratedFermions} into the common normalization factor $N$. 
In the strong coupling approximation, this reduces to 
\bea
\bra{d(p_-\,;\,\vec{0}_{\bot},\vec{d}_\perp)}\left. d(p_-\,;\,\vec{0}_{\bot},\vec{d}_\perp)
\right\rangle =\frac{1}{N}\,F_1(p_-^S),
\eea
where $F_1(p_-^S)$ is given by
\be
F_1(p_-^S)=\sum_{x_f^-,x_f^{-\,\prime}}
e^{-\mathrm{i}\,p_-^S/2 \big(x_f^--x_f^\prime{}^-\big)}
\Big\langle\onehalf\Tr{U_{-k}^f} \Big\rangle^{|x_f^--x_f^{-\,\prime}|}\,.
\ee

In the gluon correlation function one has to take care of the
arrangement of the operators.
The transversal chromo-electric field operators in the point split operator \eqnref{eqn:NLCCorrFuncLattFine} 
do not commute with the
transversal link operators appearing in the definition of the dipole operator.  
As in the previous \secref{sec:NLCCorrLat}, the string
$S_{q\bar{q}}^{f\,\dagger}[x_f^{\prime\,-}]$ arising from
the dipole at $x_f^\prime{}^-$ must appear to the left of the
point split operator $\widetilde{G}(z^-,\vec{z}_\perp\,;\,z^-{}^\prime,\vec{z}_\perp)$ 
and correspondingly the string $S_{q\bar{q}}^f[x_f^-]$ 
to the right of $\widetilde{G}(z^-,\vec{z}_\perp\,;\,z^-{}^\prime,\vec{z}_\perp)$ 
(see \figref{fig:EVOPD}). 
Then the forward matrix element of $\widetilde{G}(z^-,\vec{z}_\perp\,;\,z^-{}^\prime,\vec{z}_\perp)$ is given by 
\bea
\lefteqn{
\bra{d(p_-\,;\,\vec{0}_{\bot},\vec{d}_\perp)}\,
\widetilde{G}(z^-,\vec{z}_\perp\,;\,z^-{}^\prime,\vec{z}_\perp)\,\ket{d(p_-\,;\,\vec{0}_{\bot},\vec{d}_\perp))}
=\frac{1}{N}
\sum_{x_f^-\,,\,x_f^\prime{}^-} 
e^{-\mathrm{i}p_-^S/2\big(x_f^--x_f^\prime{}^-\big)}}
\nn\\
& & ~~~~~~~~~~~~~~~
\,\bra{\Psi_0}\Big[ S_-^f(x_f^-,\vec{d}_\perp\,;\,x_f^{\prime}{}^-;\vec{d}_\perp)\,
S_{q\bar{q}}^{f\,\dagger}\big(x_f^\prime{}^-,\vec{0}_\perp\,;\,x_f^\prime{}^-,\vec{d}_\perp\big)[x_f^\prime{}^-]\, 
S_-^f(x^\prime{}^-,\vec{0}_\perp\,;\,x^-,\vec{0}_\perp)
\Big]_{ab}
\nn\\
& & ~~~~~~~~~~~~~~~~~~~~~~~~~~~~~
\widetilde{G}(z^-,\vec{z}_\perp\,;\,z^-{}^\prime,\vec{z}_\perp)
 \Big[ S_{q\bar{q}}^f\big(x_f^-,\vec{0}_\perp\,;\,x_f^-,\vec{d}_\perp\big)[x_f^-]
\Big]_{ba}  
\ket{\Psi_0}\,.
\label{eqn:corrfunc}
\eea

The square brackets denote matrix elements with color indices $ab$ and
$ba$ respectively such that the expectation value is given by the
trace over the product of the color dipole states $S^\dagger_{q\bar{q}}$ and
$S_{q\bar{q}}^f$ with the effective point split operator
$\widetilde{G}$ in between.  
The momentum
correlation function \eqnref{eqn:corrfunc} evaluates the cross product of electric and
magnetic field strengths separated along the light cone, i.e. 
it determines the correlation of an electric field in the dipole with the
corresponding magnetic field. 
In order to compute it, we arrange the operator $\widetilde{G}$ with $\ket{\Psi_0}$ (c.f. \eqnref{eqn:VariationlGSWF}) in a way such that $\widetilde{G}$ 
stands directly in front of the trivial ground state $\ket{0}$
\bea
\widetilde{G}\,S_{q\bar{q}}^f\,\ket{\Psi_0}=
\left[\widetilde{G},S_{q\bar{q}}^f\right]\,\ket{\Psi_0}+S_{q\bar{q}}^f\,\left[\widetilde{G},\Psi_0 \right]\,\ket{0}+S_{q\bar{q}}^f\,\Psi_0\,\widetilde{G}\,\ket{0}\,.
\eea
The trivial ground state is annihilated 
by this operator. The commutator of $\widetilde{G}$ with the ground state wave
functional leaves the dipole operator intact and yields a vacuum
transition which is subtracted 
when the connected matrix element is extracted. 
Therefore, the only remaining contribution comes from the
commutator of $\widetilde{G}$ with the transversal link of the incoming dipole. 
It is given by
\bea
\lefteqn{
\left[G^f(z_f^-,\vec{z}_\perp\,;\,z_f^{-\,\prime},\vec{z}_\perp)\,,\, U_j^f(x_f^-,\vec{x}_\perp) \right]=}\nn\\
& &
\phantom{+}
\onehalf
S_-^f(x_f^{-},\vec{x}_\perp\,;\,z_f^-,\vec{x}_\perp)\,\Imag{\bar{U}_{-j}^f(z_f^-,\vec{x}_\perp)}
\,S_{-}^f(z_f^-,\vec{x}_\perp\,;\,x_f^{-},\vec{x}_\perp)
U_j^f(x^{-}_f,\vec{x}_\perp)
\,\delta_{x_f^-,z_f^{-\,\prime}}\,\delta_{\vec{x}_\perp,\vec{z}_\perp}\nn\\
& & + \,(z_f^-\leftrightarrow z_f^{-\,\prime})~.  
\eea 
Due to the interchange symmetry $z^-\leftrightarrow z^\prime{}^-$ of the commutator, only the $\cos$-part of the Fourier transformation survives:
\bea 
\lefteqn{
\sum_{z^-,z^{-}{}^\prime=-N_-/2}^{N_-/2-1}\,
e^{-\mathrm{i}\,p_-^{g}(z^--z^\prime{}^-)}
\left[\widetilde{G}(z^-,\vec{z}_\perp\,;\,z^{-\,\prime},\vec{z}_\perp)\,,\,
U_j^f(x_f^-\vec{x}_\perp) \right]}\nn\\ &=&
\sum_{z^-=-N_-/2}^{N_-/2-1}2\,\cos\left(p_-^{g}\,z^-
\right)\,S_-^f(x^-,\vec{x}_\perp\,;\,x_f^-+2\,z^-,\vec{x}_\perp)\,\Imag{\bar{U}_{-j}^f(x_f^-+2\,z^-,\vec{x}_\perp)}\nn\\
& & ~~~~~~~~~~~~~~~~~~~\times S_{-}^f(x_f^-+2\,z^-,\vec{x}_\perp\,;\,x_f^{-},\vec{x}_\perp)\,U_j^f(x_f^{-},\vec{x}_\perp)
\,\delta_{\vec{x}_\perp,\vec{z}_\perp}\,.  
\eea

The gluon distribution function \eqnref{eqn:GluonDistLatCoarse} for a 
one-link dipole with total
string momentum $p_-^S$ becomes
\bea
g_1(p_-^{g};p_-^{S})&=&\frac{4}{\pi}\frac{1}{p_-^{g}}\sum_{z^-}
\cos\left(p_-^{g}\,z^- \right)
\frac{N_-}{F_1(p_-^{S})}\sum_{x_f^-}e^{\mathrm{i}\,p_-^S/2\,x_f^-}\nn\\ &
&\bra{\Psi_0}\mathrm{Tr}
\left[S_-^f(0,\vec{d}_\perp\,;\,x_f^-,\vec{d}_\perp)\,
U_j^{f\,\dagger}(x_f^-,\vec{0}_\perp)\,S_-^f(x_f^-,\vec{0}_\perp\,;\,2\,z^-,\vec{0}_\perp)\right.\nn\\
& &~~~~~~\left. \times
\Imag{\bar{U}_{-j}^f(2\,z^-,\vec{0}_\perp)}\,S_{-}^f(2\,z^-,\vec{0}_\perp\,;\,0,\vec{0}_\perp)\,U_j^f(0,\vec{0}_\perp)\right]\ket{\Psi_0}\,.
\label{eqn:NLCCorrFuncUnevaluated}
\eea
In $g_1(p_-^{g};p_-^{S})$
we indicate the total string momentum by the argument $p_-^S$ and the number
of transversal links by the index $n=1$.  

\begin{figure}[h]
\centering
\includegraphics[width=.8\textwidth]{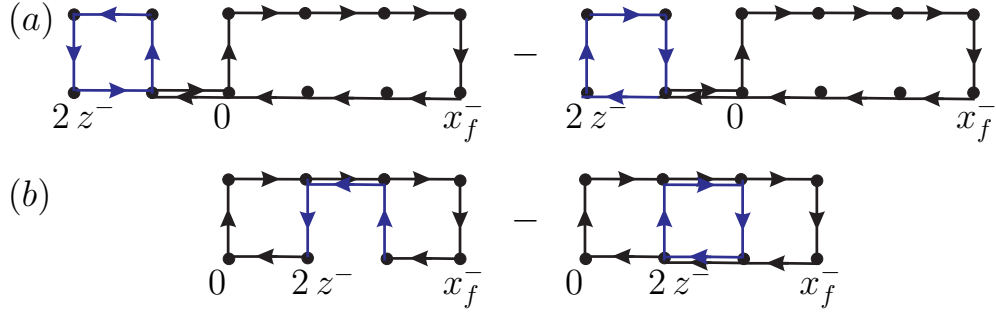}
\caption{\label{fig:PlotOLExpectVal} 
Visualization of the matrix element in \eqnref{eqn:NLCCorrFuncUnevaluated}. 
There are
two different possible situations resulting in different expectation values of the operator, 
namely either the field strength lies outside of the Wegner-Wilson loop (a) or it lies inside (b).
The difference in each of the two cases corresponds to the 
antihermitean part of the plaquette. }
\end{figure}

In the following, we discuss the evaluation of the matrix element
in \eqnref{eqn:NLCCorrFuncUnevaluated}.
The imaginary part of a $U_{-j}^f$ plaquette (field strength) located at longitudinal position $2 \, z^-$
has to be correlated with a closed loop of links in the longitudinal
transversal plane located between longitudinal positions $0$ and $x^-_f$ 
as visualized in \figref{fig:PlotOLExpectVal}. 
The field strength is connected with the edge of the Wegner-Wilson 
loop at longitudinal position $0$. 
We distinguish two cases depicted in \figref{fig:PlotOLExpectVal} which can be fully 
evaluated in the strong coupling approximation: 
\begin{itemize}
\item[(a)] The plaquette $U_{-j}^f$ lies outside of the loop. The 
           matrix element factorizes 
           and vanishes due to $\mathrm{Tr}[\mathrm{Im} (U_{-j}^f)]=0$.
\item[(b)] The plaquette lies inside the loop, the matrix element 
           can be computed by tiling which yields
           the second line of \eqnref{eqn:GluonDistSingleLink}.           
\end{itemize}
One finally obtains for the gluon distribution function of a one-link dipole
\bea
g_1(p_-^{g},p_-^{S})&=&\frac{4}{\pi}\frac{1}{p_-^{g}}\sum_{z^-}\cos\left(p_-^{g}\,z^-
\right)
\frac{N_-}{F_1(p_-^{S})}\sum_{x_f^-} \sin\left(\frac{p_-^S}{2}\,x_f^-\right)\nn\\
& & \left(1-\Big\langle \Big(\onehalf\Tr{U_{-k}^f}\Big)^2 \Big\rangle
\right) \Big\langle\onehalf\Tr{U_{-k}^f} \Big\rangle^{|x_f^-|-1}\nn\\ 
& & \times \left[ \onehalf \Big(
\Theta_0(2\,z^-)\Theta(x_f^--2\,z^-)+\Theta(2\,z^-)\Theta_0(x_f^--2\,z^-)
\Big)\right.\nn\\ & &~~~~~~~~\left.- \onehalf \Big(
\Theta(-2\,z^-)\Theta_0(2\,z^--x_f^-)+\Theta_0(-2\,z^-)\Theta(2\,z^--x_f^-)
\Big) \right]\,.  
\label{eqn:GluonDistSingleLink}
\eea
The two cases a) and b) are encoded in the $\Theta$-distributions $\Theta_0(x)$ and $\Theta(x)$ differing
in the value they take at $x=0$, i.e.
\be
\Theta(x)\equiv\left\{
\begin{array}{lcl}
1 & , & x>0 \\
0 & , & x\leq 0
\end{array}
\right.
~~,~~
\Theta_0(x)\equiv\left\{
\begin{array}{lcl}
1 & , & x\geq 0 \\
0 & , & x < 0
\end{array}
\right.
~.
\ee

In the first line of \Eqnref{eqn:GluonDistSingleLink},  
only the imaginary part $\sin(p_-^S/2\,x_f^-)$ survives
from the exponential $e^{-\mathrm{i}\,p_-^S/2\,x_f^-}$ due to the antisymmetry of the sum in $x^-_f$.  

Let us now check the gluon momentum sum rule
by evaluating the expectation value of the longitudinal momentum operator $\langle p_-^g\rangle=\sum_{z^-,\vec{z}_\perp} \left\langle \widetilde{\mathcal{P}}_-(z^-,\vec{z}_\perp) \right\rangle$ 
as a function of the external string momentum
$p_-^S$.
\begin{figure}
\centering
\includegraphics[width=.8\textwidth]{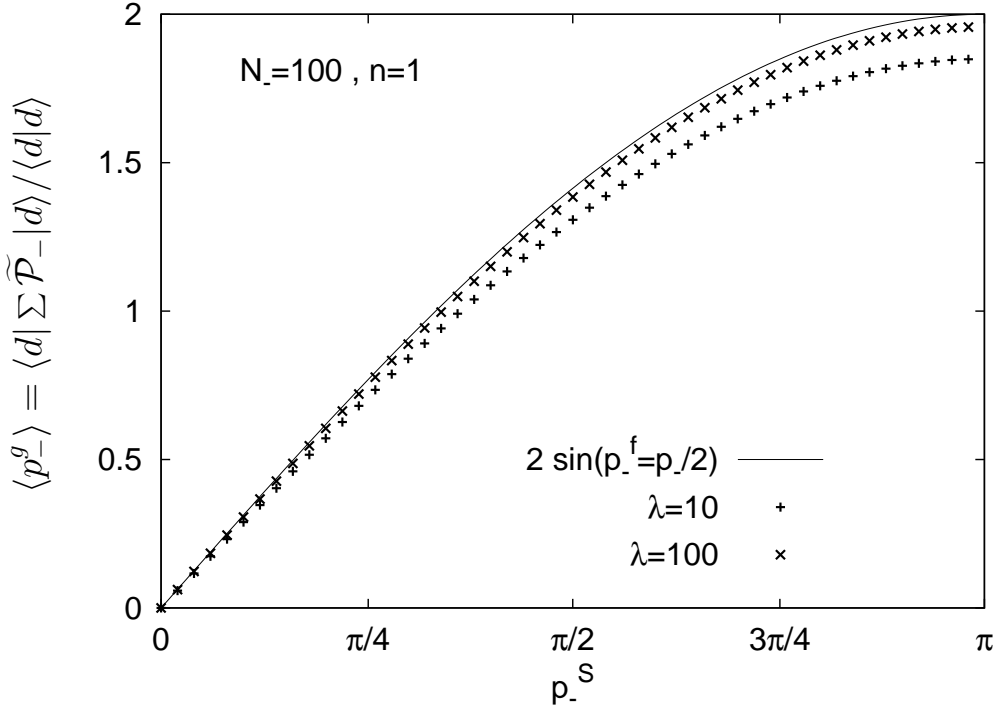}
\caption{\label{fig:MomentumConservation} 
With crosses we show the expectation value of the longitudinal momentum operator $\widetilde{P}_-$
of a one-link dipole state as a function of $p_-=p_-^S$ on a $N_-=100$ lattice for $\lambda=10$ and $100$.
We also show the expected eigenvalue $2\,\sin(p_-^f=p_-^S/2)$ 
of the longitudinal momentum operator without $\mathcal{O}(a^2)$ 
corrections of a single-link color dipole state projected onto 
longitudinal momentum $p_-$.    }
\end{figure}
In \figref{fig:MomentumConservation} we show the 
mean gluon momentum as a function of the lattice momentum $p_-^S$. 
The expectation value is obtained on a $N_-=100$ lattice for 
two different values of $\lambda$, i.e. $\lambda=10$ and $100$. 
In the region of small momenta $p_-^S$
one can recognise that the lattice discretisation is very accurate. Due to the introduction of the finer sublattice the mapping of the
lattice momentum $p_-^S$ to the mean momentum $\langle p_-^g \rangle$ is unique. 
For comparison we also show the exact eigenvalue $2 \sin(p_-^S/2)$ without order $\mathcal{O}(a^2)$
corrections.
The larger the lattice coupling constant, the more accurate becomes 
the mapping from lattice momenta to observed momenta.

\begin{figure}[h]
\centering
\includegraphics[width=.8\textwidth]{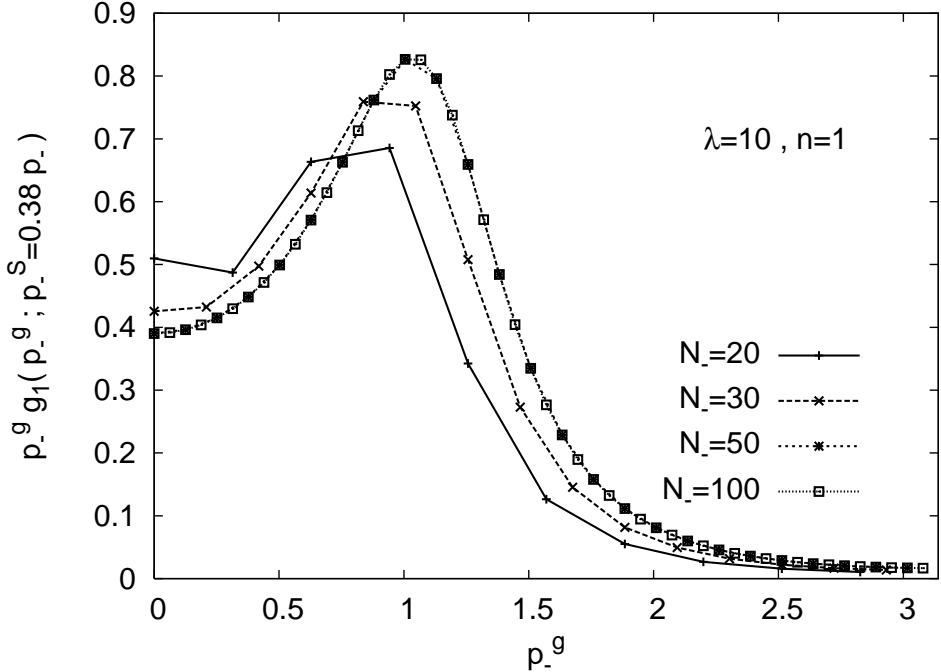}
\caption{\label{fig:PlotScaling} 
Gluon distribution function $p_-^g\,g_1(p_-^g\,;\,p_-^S=0.38\,p_-)$ of a color dipole 
with a single transversal link ($n=1$)
in the transversal direction for different lattice sizes $N_-$. The distribution functions are 
computed with a lattice coupling $\lambda=4/g^4=10$ in the effective $SU(2)$ lattice Hamiltonian
of \eqnref{eqn:EffectiveLatHamiltonian}. The average gluon momentum $\langle p_-^g \rangle =p_-^S$ 
of the dipole state has been adjusted to the average gluon momentum $\langle x_B\rangle\,p_-=0.38\,p_-$ of the \emph{MRST} gluon distribution function at $Q^2=1.5\,\mathrm{GeV}^2$.  
}
\end{figure}

The full gluon distribution function
$p_-^g\,g_1(p_-^g;p_-^S)$ of a color dipole with one link in the
transversal direction is shown in \figref{fig:PlotScaling}.  
It is computed on lattices
with $N_-=20$, $30$, $50$, and $100$ sites in the longitudinal direction.
The average gluon
momentum $\langle p_-^g\rangle$ of the dipole state has been adjusted to the average
gluon momentum $\langle p_-^g\rangle=0.38\,p_-$ of the \emph{MRST}
gluon distribution function at $Q^2=1.5\,\mathrm{GeV}^2$. 
So far,
the simulated gluon distribution function on the lattice depends on the 
total hadron momentum. This is in contrast to Feynman scaling,
where the gluon distribution function 
is only a function of the gluon momentum fraction 
$ p_-^g /p_-$.
For fixed $\lambda$, the hadronic lattice momentum $p_-=2\,\pi\,(N_-/2-1)/N_-$ is exclusively determined by the longitudinal lattice extension $N_-$. 
Independence
of the gluon distribution function on the hadronic momentum would be equivalent to independence 
on the lattice extension.
This figure demonstrates the effect of 
increasing the number of longitudinal lattice sites, i.e. approaching the infinite volume limit.
Scaling for $\lambda=10$ seems to be obeyed for longitudinal lattice extensions larger than $N_-=50$.  
Realistic lattice simulations with an improved ground state wave functional need quite large longitudinal lattice sizes.
The smearing of the distribution function 
is due to the gluon dynamics incorporated in the Wegner-Wilson loop 
expectation value.
The area law behavior of the Wegner-Wilson loop yields a non-trivial gluon wave 
function which broadens the distribution.

\begin{figure}[h]
\centering
\includegraphics[width=.8\textwidth]{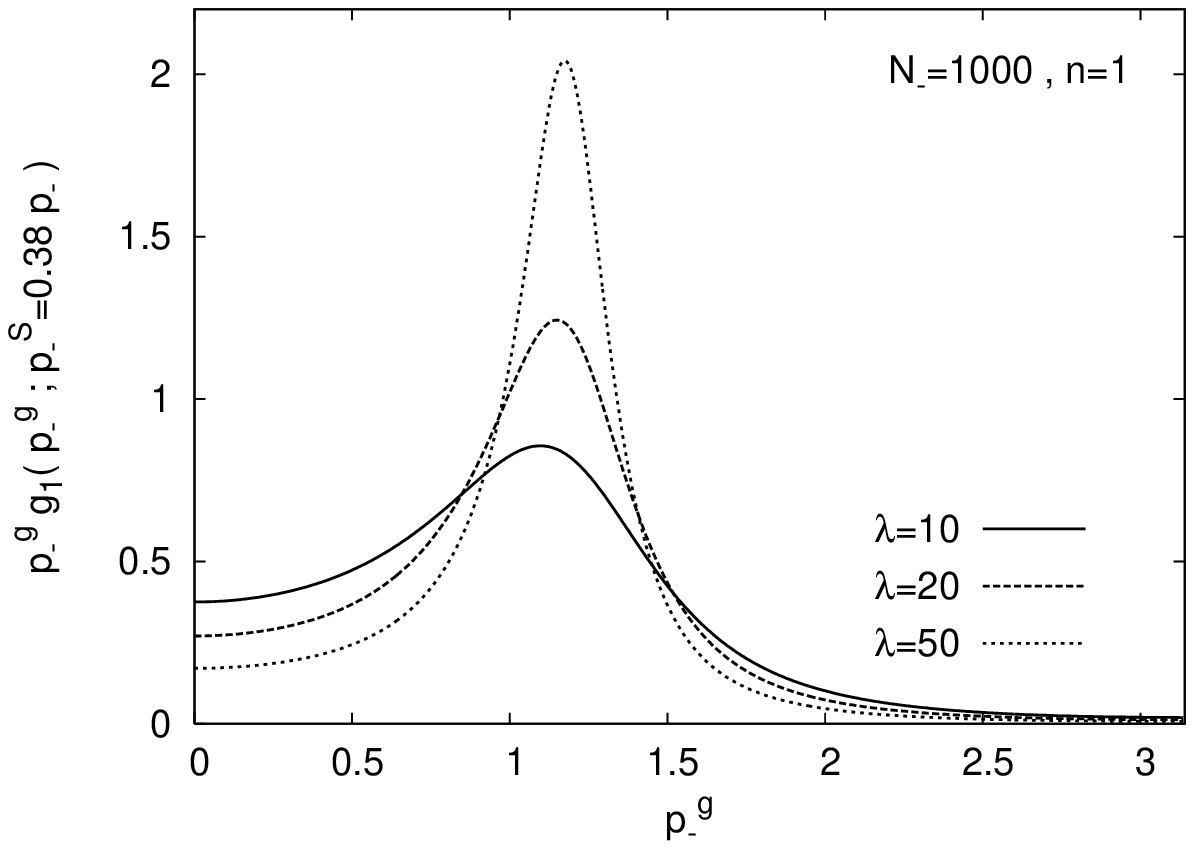}
\caption{\label{fig:GluonDistrLambda} 
Gluon distribution function $p_-^g\,g_1(p_-^g ; p_-^S=0.38\,p_-)$ of a color dipole with a single transversal link ($n=1$)
in the transversal direction at different lattice gauge couplings $\lambda$. The distribution functions have been computed on a lattice with $N_-=1000$ sites in the longitudinal direction. The average gluon momentum $\langle p_-^g \rangle$ of the dipole state has been adjusted to the average gluon momentum $\langle x_B\rangle \,p_-=0.38\,p_-$ of the \emph{MRST} gluon distribution function at $Q^2=1.5\,\mathrm{GeV}^2$ .  }
\end{figure}

If one varies the lattice gauge coupling
$\lambda$, the single plaquette expectation values vary between $0$ and $1$ 
depending on the coupling constant $\lambda$. This has consequences for 
the width of the gluon distribution function
$p_-^g\,g_1(p_-^g;p_-^S)$ as shown in \figref{fig:GluonDistrLambda}.
The larger $\lambda$, 
i.e. the smaller the QCD gauge coupling
$g^2$, 
the stronger the peak in the one-link distribution function becomes.

In the extreme weak coupling limit $\lambda\rightarrow\infty$ the single
plaquette expectation value approaches $1$. For finite coupling constants, the single 
plaquette expectation value is less than one. 
Hence, it suppresses large Wegner-Wilson loop 
extensions in \eqnref{eqn:GluonDistSingleLink}:
\be
W(0,\vec{0}_\perp\,;\,x_f{}^-,\vec{d}_\perp)=\Big\langle \onehalf\, \Tr{ U_{-k}^f } \Big\rangle^{|x_f^-|}\,.
\ee
One can define a correlation length which represents
the longitudinal distance 
at which the single plaquette expectation value reduces to one half of its original value 
\be
\Delta \xi=\log(\onehalf)/\log \left(\Big\langle \onehalf\, \Tr{ U_{-k}^f } \Big\rangle\right).
\label{eqn:DeltaXi}
\ee
By using \eqnref{eqn:OptParam}, \eqnref{eqn:StrongCouplingApproxII} and \eqnref{eqn:DeltaXi}, we can evaluate the correlation length directly as a function of $\lambda$.
At $\lambda=10$, the correlation length is given by
$\left.\Delta \xi\right|_{\lambda=10}=3.12$ and at $\lambda=50$ it is given by $\left.\Delta \xi\right|_{\lambda=50}=8.05$.
Since the nlc gluon correlation function only has support when it lies inside the Wegner-Wilson loop,
the correlation length $\Delta \xi$ gives an estimate for 
the width $\Delta p_-^g$ of the gluon momentum distribution,
\be
\Delta p_-^g=\frac{1}{\Delta\xi}\,.
\ee
This implies that the width of the gluon momentum distribution at $\lambda=10$ is given
by $\left.\Delta p_-^g\right|_{\lambda=10}=0.32$ and at $\lambda=50$ by $\left.\Delta p_-^g\right|_{\lambda=10}=0.13$ as seen in \figref{fig:GluonDistrLambda}.
In the extreme weak coupling limit, when the link reduces to a single gluon, the gluon distribution function is sharp, i.e. $g_1(p_-^g;p_-^S)=\delta(p_-^g-p_-^S)$.  Since the link momentum is fixed by the projection
onto a definite momentum, also the gluon momentum is fixed in this limit and no variance 
is allowed. On the other hand, for smaller values of $\lambda=\frac{4}{g^4}$, 
i.e. for strong coupling the correlation length becomes smaller 
which implies that one has a broad momentum distribution
peaked around $p_-^S$. 
   
\section{The gluon distribution function of a hadron}
\label{sec:HadronicGluonDist}

Up to now, we have considered the gluon distribution function of a
color dipole consisting of a single transversal link.  The one-link
dipole gluon distribution is the basic building block from which the
multiple link dipole gluon distribution function of a hadron can be
constructed.  We expand a hadronic state 
$\ket{h(p_-,\vec{0}_\perp)}$ in dipole components,
i.e.  
\bea
\ket{h(p_-,\vec{0}_\perp)}&=&
\sum_{\mathcal{C},\vec{d}_\perp}
\Psi_h(\mathcal{C},\vec{d}_\perp)\ket{d(p_-\,;\,-\vec{d}_\perp/2,\mathcal{C}_\perp,\vec{d}_\perp/2)}\,,
\nn\\
\Psi_h(\mathcal{C},\vec{d}_\perp)&\equiv& \Big\langle
d(p_-\,;\,-\vec{d}_\perp/2,\mathcal{C}_\perp,\vec{d}_\perp/2) \Big.\Big|
h(p_-,\vec{0}_\perp)\Big\rangle\,.  
\eea 

The hadron has the same momentum $p_-$ and the same transversal cm coordinate $\vec{x}_\perp=0$ 
as the dipoles.
The wave function  $\Psi_h(\mathcal{C}_\perp,\vec{d}_\perp)$ 
represents the probability amplitude  to find a dipole
with a quark and an antiquark separated by the transversal distance $\vec{d}_\perp$ and 
connected by the path $\mathcal{C}_\perp$ in the hadron.  Hence, the actual hadronic gluon
distribution function arises from a superposition of multiple link configurations.
The wiggly strings $S_{q\,\bar{q}}$ c.f. \eqnref{eqn:FinalDipoleState} 
connecting the quark/antiquark are not restricted to lie along one of the coordinate axes
(c.f. \figref{fig:Dipole}), but have a fixed 
number of transversal links as explained in \secref{sec:DipoleModel}.
In order to project
this state on angular momentum $J_z=0$, we rotate the 
hadron in the transversal plane by summing over randomly chosen curves $\mathcal{C}_\perp$
which can be constructed in the following way.  A random walker
starts at an initial time $t=0$ and trails a Schwinger string along
its path through the transversal lattice.  In each time step, the walker may hop
with equal weight in one of the four transverse directions.
The random walk ends if the number of hops corresponds to the number
of allowed transversal links fixed by the energy constraint.  The
starting point of the random walker has to be chosen a posteriori in
such a way that the center of mass of the generated dipole
configuration is at the origin.  The ensemble of possible random paths
automatically obeys the desired rotational symmetry.

Since the energy of the strings with a given number of transversal links is the 
same for all the string configurations, we assume that the probabilities
among the total number  $\#$ of curves with n-links
are equally distributed:
\be
|\Psi_h(\mathcal{C},\vec{d})|^2=\frac{1}{\#}\,.
\ee

From the random walk follows that for n-links  
the hadron has an average radius squared $\vec{R}_\perp^2$ proportional to n: 
Hence, the area of the hadron scales with the number of links
\be
\left\langle R_\perp^2 \right\rangle = \frac{n\,a_\perp^2}{2}~.
\ee

Due to the strong coupling approximation non vanishing gluonic matrix elements 
need incoming and outgoing states to have the same curve connecting the 
quark and the antiquark:
\bea
\lefteqn{ 
\bra{h(p_-,\vec{0}_\perp)}O\ket{h(p_-,\vec{0}_\perp)}=}\nn\\
& &
\sum_{\mathcal{C},\vec{d}_\perp}
\,\left|\Psi_h(\mathcal{C},\vec{d}_\perp)\right|^2\,\bra{d(p_-\,;\,-\vec{d}_\perp/2,\mathcal{C}_\perp,\vec{d}_\perp/2)}O\ket{d(p_-\,;\,-\vec{d}_\perp/2,\mathcal{C}_\perp,\vec{d}_\perp/2)}\,.
\eea

Because of the equal weight of all the dipole configurations with fixed transversal length n, 
the gluon  distribution can be calculated from the distribution function $g_n$ 
of a string elongated along only one of the transversal axes 
(c.f. \figref{fig:Dipole}):
\be
g_h(p_-^g;p_-^S)=g_n(p_-^g;p_-^S) \, .
\ee

Due to the sum rule (c.f. \eqnref{eqn:MomSumRuleLat}) the expectation value of the 
gluon momentum inside the $n$-link dipole is fixed as $\langle p_-^g \rangle=p_-^S$.     
The computation of $g_n$ is done in analogy to the 
computation of the single-link gluon distribution. 
Because of 
the summation of the nlc correlation function over the entire transversal lattice, 
the chromo-electric field $\Pi_k^a$ in the operator defining the 
nlc correlation function, can act on each of the
transversal links appearing in the connector $S_{q\bar{q}}$. In
strong coupling the total loop factorizes,
therefore  the $n$-link distribution function is given  by 
the product of a splitting function $P_{n\rightarrow n-1}$ 
multiplying the gluon
distribution function with $n-1$ links . In this 
recursion relation (c.f. \appref{app:Relgn2g1}) all possible
intermediate momenta of the substring are summed over: 
\be 
g_n\big(p_-^g;p_-^S \big)=
\frac{2\,\pi}{N_-}\sum_{p_-^S{}^\prime=0}^{p_-^S}g_{n-1}\big(p_-^g;p_-^S{}^\prime\big)\,P_{n\rightarrow
n-1}(p_-^S,p_-^S{}^\prime) \, .
\label{eqn:DipoleRecursionRel}
\ee

The splitting function  $P_{n\rightarrow n-1}(p_-^S,p_-^S{}^\prime)$
denotes  the probability that a string
with $n$ transversal links and total momentum
$p_-^S$ splits into a string with $n-1$ transversal links
and total momentum $p_-^S{}^\prime$. The form of the 
splitting function is of a kinematical and a dynamical origin which are both encoded in
the functions $F_{m}(p_-^S)$ for $m=n-1$ and $m=1$ due to the recursive 
representation:
\be 
P_{n\rightarrow n-1}(p_-^S,p_-^S{}^\prime)=
\left(\frac{n}{n-1}\right) \times \frac{F_{n-1}\big(p_-^S{}^\prime\big)\,F_{1}
\big(p_-^S-p_-^S{}^\prime\big)}{\frac{2\,\pi}{N_-}\sum_{p_-^S{}^{\prime\prime}=0}^{p_-^S}F_{n-1}
\big(p_-^S{}^{\prime\prime}\big)\,F_{1}\big(p_-^S-p_-^S{}^{\prime\prime}\big)}\,.
\label{eqn:DipoleRecursionRel1}
\ee

The functions $F_{n-1}(p_-^S)$ and $F_1(p_-^S)$     
contain momentum conservation in the density matrix $\rho_n$,
\be
\rho_n[p_-^S,y_j^-,y_j^\prime{}^- ]=\sum_{l_-^j,l_-^{j\prime}}\,
\delta(p_-^S-\sum_{j=1}^n\,l_-^{j})
\,\delta(p_-^S-\sum_{j=1}^n\,l_-^{j\,\prime})\,
e^{-\mathrm{i}\,\sum\limits_{j=1}^n
\big(l_-^{j}/2\,y_j^--l_-^{j\,\prime}/2\,y_j^\prime{}^-\big)}\, ,
\label{eqn:RhoDef}
\ee
and the gluon dynamics in the residual part of $F_n(p_-^S)$ is related to the
n-fold product of expectation values of Wilson loops with one link in
transverse direction and $|y^--y^{'-}|$ links in longitudinal
direction (cf. \eqnref{eqn:NormNLinkDipole}):
\be
F_n(p_-^S)=\sum_{\{y_j^-\},\{y_j^-{}^\prime\}}\,\rho_n[p_-^S;y_j^-,y_j^-{}^\prime]\,
\prod_{j=1}^n\Big\langle\onehalf\Tr{U_{-k}}
\Big\rangle^{|y_j^--y_j^-{}^\prime|} \, .
\label{eqn:NormNLinkDipole}
\ee 

The denominator of \eqnref{eqn:DipoleRecursionRel1} guarantees the
correct normalization of the splitting function $P_{n\rightarrow
n-1}(p_-^S,p_-^S{}^\prime)$ which has to satisfy the following
relation
\be 
\frac{2\,\pi}{N_-}\sum_{p_-^S{}^\prime=0}^{p_-^S} p_-^S{}^\prime \,P_{n\rightarrow
n-1}(p_-^S,p_-^S{}^\prime)=p_-^S \, , 
\ee
in order that the distribution function $g_n(p_-^g,p_-^S)$ obeys the momentum sum rule
\be
\frac{2\,\pi}{N_-}\sum_{p_-^g=0}^{p_-} p_-^g \, g_n(p_-^g;p_-^S)=p_-^S \, .
\ee
  
The initial condition  for 
the recursion relation \eqnref{eqn:DipoleRecursionRel} is given 
the one-link dipole function $g_1(p_-^g;p_-^S)$ derived in \eqnref{eqn:GluonDistSingleLink}
with the total gluon momentum fraction taken from experiment.
We use as lattice gauge coupling $\lambda=4/g^4=10$ unless otherwise
noted which corresponds to typically ``strong coupling `` transverse 
lattice size far from the continuum $a_\perp=0.5-0.6 fm$, 
i.e. to an input scale of
$Q^2\approx\pi^2/a_\perp^2=1.5 \,\mathrm{GeV}^2$. One can try to devolve the
phenomenological \emph{NLO MRST 2002} \cite{Martin:2001es} and the \emph{CTEQ 6AB} parameterizations \cite{Pumplin:2005rh}
of the gluon distribution function and one finds $p_-^S=0.38\, p_-$. 
Since the computation is purely arithmetic in strong coupling , we can 
use a large longitudinal lattice with $N_-=1000$ lattice sites.
The so defined lattice gluon distribution function depends on the gluon
lattice momenta $p_-^g= 2\, \pi \,n / N_-$ where $N_-$ is the number of
lattice sites in longitudinal direction on the coarse lattice and the
integer $0 \leq n \leq N_-/2-1$. With
$N_-=1000 $ we find  a smooth limit for the structure function, which
we can associate naively as a  scaling structure function 
(cf. \figref{fig:PlotScaling}).
The simple vacuum wave functional we use does not allow us to discuss 
the continuum light cone limit with the longitudinal lattice size $N_-\,
a_-$ constant when $N_- \rightarrow \infty$ and $a_-
\rightarrow 0$. It has been shown in the Schwinger model \cite{Lenz:1991sa,Vary:1996uc} 
that the infinite volume limit has to be performed before
the light cone limit.

\begin{figure}[h]
\centering
\includegraphics[width=.8\textwidth]{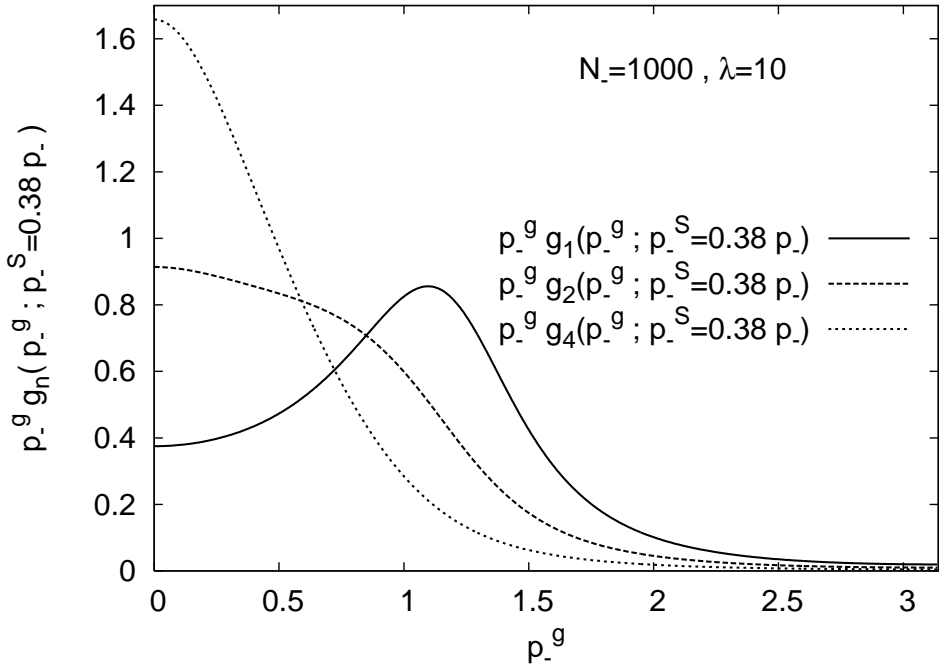}
\caption{\label{fig:GluonDistr} 
Gluon distribution function $x_B\,g_n(x_B;p_-^g)$ of a color dipole with different number of links ($n=1,2,4$) in the transversal direction. The distribution functions have been computed on a lattice with $N_-=1000$ sites in the longitudinal direction at a lattice gauge coupling $\lambda=10$. The average gluon momentum $p_-^g$ of the dipole state has been adjusted to the average gluon momentum $\langle x_B\rangle=0.38$ of the \emph{MRST} gluon distribution function at $Q^2=1.5\,\mathrm{GeV}^2$.  
}
\end{figure}

If one increases the number of transversal links, the gluons have
access to a larger region in phase space due to the splitting function $P_{n\rightarrow n-1}$ 
in \eqnref{eqn:DipoleRecursionRel}. An  increase in the number of
transversal link operators implies that the total gluon momentum will
be partitioned among more gluons.  Hence, it becomes more likely to find
a gluon with a small fraction of the total momentum. This can be
observed in \figref{fig:GluonDistr}.  The mean momentum fraction of the 
gluons i.e. the integral under the curve
remains constant, however, the gluons with large momenta
are shifted from large to smaller values of $p_{-}^g$.

We can discuss what happens when one
increases the resolution. Keeping the transversal extension of the dipole fixed 
we increase the number of transversal links by one unit, then the recursion relation
\eqnref{eqn:DipoleRecursionRel} gives a strong coupling
equivalent to the weak coupling DGLAP equation, which describes the
change of the parton distribution function under a variation of 
resolution. Indeed, if one has scaling in the limit $p_-^S
\rightarrow\infty$ the recursion relation
\eqnref{eqn:DipoleRecursionRel} can be written as
\bea
g_n(x_B) &=& \int_{x_B}^{1} \frac{dz_B}{z_B}\,g_{n-1}(x_B/z_B)\,P_{n\rightarrow n-1}(z_B) \, , \nonumber \\
x_B     &=&p_-^g/p_-^S \, , \nonumber \\
z_B     &=&p_-^S{}^\prime/p_-^S \, .
\eea

Subtracting $g_{n-1}$ from $g_n$  one arrives at an equation which has almost 
the form of the weak coupling DGLAP equation. 
The main difference occurs in the redefined splitting
function $P_{n\rightarrow n-1}$. 
In the usual DGLAP equation, the splitting function denotes the probability for a 
gluon to split into two gluons, one of them carrying the momentum fraction $z_B$. 
Our equation resembles more the LUND model \cite{Andersson:1983ia} where the 
dynamics of the entire fragmenting string is described. 
One can see that in the weak coupling limit the plaquette expectation values become unity 
plus $O(g^2)$ corrections and make the redefined splitting function $P-1$ proportional to 
$\alpha_s$.
Once the continuum limit is under control
with a suitable wave functional of the ground state, one may consider
the transition of the so redefined splitting function into the DGLAP Kernel. 

\begin{figure}[h]
\centering
\includegraphics[width=.8\textwidth]{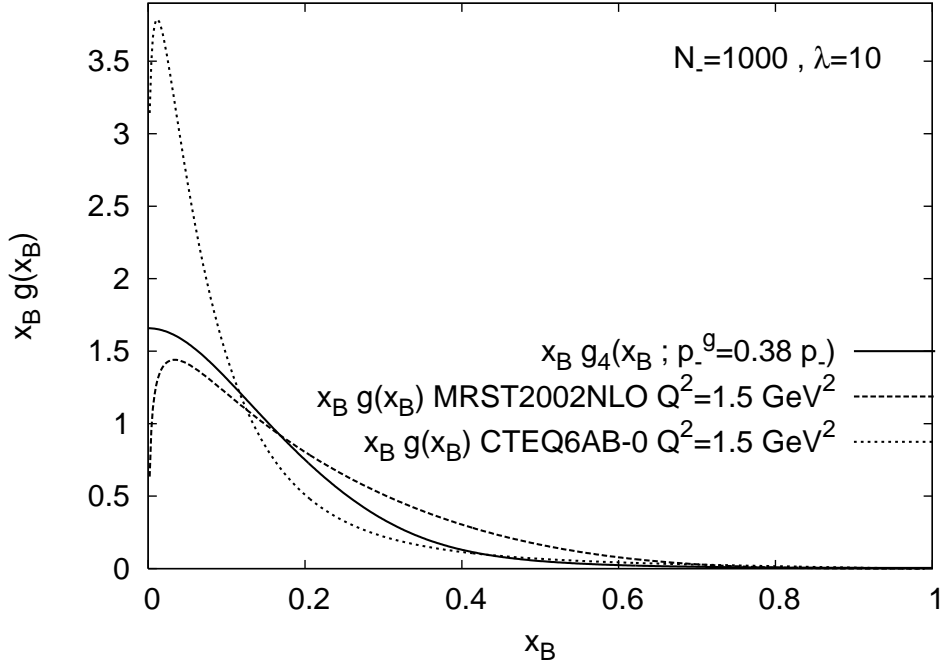}
\caption{\label{fig:GluonDistrComp} Gluon distribution function
$x_B\,g_n(x_B;p_-^g)$ of a color dipole whose number of links in the
transversal direction is given by $n=4$ in comparison
with the \emph{MRST} and the \emph{CTEQ} gluon distribution
function at $Q^2=1.5\,\mathrm{GeV}^2$.  The lattice distribution
function has been computed on a lattice with $N_-=1000$ sites in the
longitudinal direction at a lattice gauge coupling $\lambda=10$. The
average gluon momentum $p_-^g$ of the dipole state has been adjusted
to the average gluon momentum $\langle x_B\rangle=0.38$ of the
\emph{MRST} gluon distribution function at
$Q^2=1.5\,\mathrm{GeV}^2$.  }
\end{figure}

In \figref{fig:GluonDistrComp}, we compare the theoretical
gluon structure function for a
$n=4$ link dipole with the \emph{MRST} and the \emph{CTEQ} gluon
distribution function at $Q^2=1.5\,\mathrm{GeV}^2$ as  functions of
the gluon fractional momentum $x_B=p_-^g/p_-$. 
As before, the first moment of the
lattice gluon distribution function has been fixed in this figure to
the value $\langle x_B\rangle=0.38$ at $Q^2=1.5\,\mathrm{GeV}^2$. The
average gluon fractional momentum obtained from the \emph{CTEQ}
parameterization differs only by ten per cent from the
\emph{MRST} value. 
We choose four links to be consistent with the size of the proton and 
the relation $\left\langle R_\perp^2 \right\rangle = \frac{n\,a_\perp^2}{2}~$
and a transversal lattice size of $a_\perp \approx 0.65~\mathrm{fm}$.
The functional
behavior of the gluon distribution function as a function of $x_B$
multiplied with $x_B$ is the same as the functional behavior of the
gluon distribution function as a function of $p_-^g$ multiplied with
$p_-^g$, 
\be 
x_B \, g_n(x_B;p_-^S)=\left. p_-^g\,g_n(p_-^g;p_-^S)
\right|_{p_-^g=x_B\,p_-} \, .  
\ee 

The lattice gluon distribution function agrees within the systematic
uncertainty with the phenomenological \emph{MRST} -gluon distribution function .
But the figure shows that there is a large systematic uncertainty in the gluon distribution
function evolved to $Q^2=1.5\,\mathrm{GeV}^2$ depending on
the different parameterizations. 
The \emph{MRST} collaboration even gives negative values of the gluon distribution function
at small values of $x_B$ for such a small $Q^2$.

An important property of the gluon distribution function at small values of $x_B$ is 
its dependence on hadronic size. One knows hadronic cross sections at intermediate energies
and can deduce that the gluon structure function at small $x_B$  or the soft Pomeron coupling 
depends on the area of the hadron \cite{Povh:1987ju}.
With decreasing $x_B$  the gluons become  uniformly distributed inside the hadron
such that the gluon distribution function should indeed be proportional
to the transversal area $\pi \langle R_\perp^2 \rangle$ of the hadron. 

\begin{figure}[h]
\centering
\includegraphics[width=.8\textwidth]{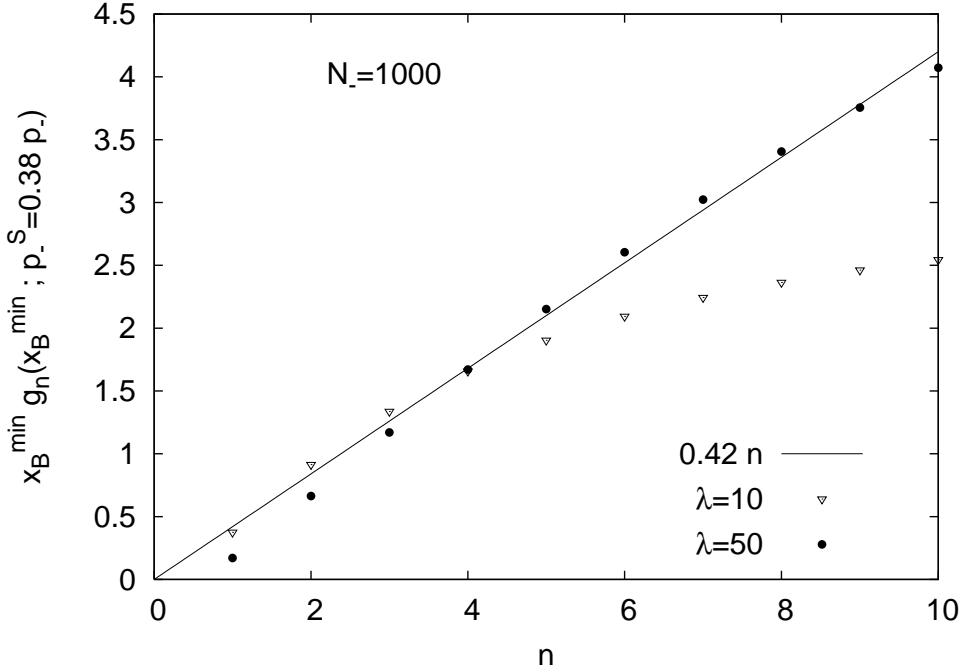}
\caption{\label{fig:Saturation} 
Gluon distribution function $x_B^{\mathrm{min}}\,g_n(x_B^{\mathrm{min}};p_-^S)$ as a function
of the number of transversal links $n$ for two different values of $\lambda$.  
The average gluon momentum $p_-^S$ of the dipole state has been adjusted to the average gluon momentum $\langle x_B\rangle=0.38$ of the \emph{MRST} gluon distribution function at $Q^2=1.5\,\mathrm{GeV}^2$.  
}
\end{figure}

In \figref{fig:Saturation}, we 
show the gluon distribution function at the lowest
value of $x_B$ compatible with the lattice momentum cut-off, i.e. $x_B=x_B^{\mathrm{min}}
=0.002$ as a function of $n$
for two different values of $\lambda$:
\be
x_B^{\mathrm{min}}=\frac{2}{N_--2} \, .
\ee

For $\lambda=50$,  
the gluon distribution function at $x_B^{\mathrm{min}}$
depends linearly on the hadronic size $n=2\,\langle R_\perp^2\rangle/a_\perp^2$, i.e. one
obtains the expected dependence of the ``hadron cross-section'' .
In order to guide the eye, we also plot the best
fit with
\be
x_B^{\mathrm{min}}
\left. g_n(x_B^{\mathrm{min}};p_-^S) \right|_{\lambda=50}=c\,n~~,~~c=0.42
\ee
into the plot. 

For $\lambda=10$, i.e. for stronger coupling , the dependence of the gluon
distribution function at $x_B^{\mathrm{min}}$ on $n$ is less then
a linear. In the strong coupling regime, rotational invariance on the lattice
is broken. Therefore, the cross section of the hadron is no longer given 
by a circular disk.

\section{Summary and outlook}
\label{sec:Summary}
\noindent

In high energy scattering
partons move along almost light like trajectories. Hence, 
light cone coordinates define the appropriate framework in order to 
describe high energy scattering experiments. If one wants to apply
the computational methods of lattice gauge theory usually defined in 
a Euclidean signature to the computation of observables of 
high energy scattering 
experiments one has to face the problem that the light cone shrinks 
to a single point. Hence, correlation functions along the light cone
which are important for the determination of structure functions can not be
computed directly. One needs the operator product expansion
in order to compute these correlations on the lattice. By doing so, one 
is restricted to the moments of the structure function. 

We have proposed to use the nlc lattice formulation in order to compute the 
correlation functions on the light cone directly. 
We have generalized the definition of the light cone correlation function
to nlc coordinates, such that in the light cone limit the original definition
is recovered and that the nlc correlation function obeys momentum conservation. 
In our approach, we are not restricted to the moments of the gluon distribution
function. 

We employ the nlc ground state wave functional in the light cone limit 
which was variationally optimized  
close to the light cone limit (cf. Ref.~\cite{Grunewald:2007cy}). Since our 
theory is formulated in a Hamiltonian framework, we stay in Minkowski 
space-time throughout the computation. This implies that one does not need 
to perform an analytical continuation from a Euclidean to a Minkowskian 
signature at the end of the computation.  

The nlc ground state wave functional ansatz in the light cone limit shows
a significant simplification for the computation of gluonic matrix elements 
in comparison to an equivalent equal time quantised computation. 
The ground state wave functional decouples the purely transversal dynamics. 
Hence, one effectively deals with 
two two-dimensional gauge theories, each living in a set of 
longitudinal-transversal planes which are distinguished by 
the other transversal coordinate. 
If one neglects boundary terms, the analytical tools for the computation 
of matrix elements valid in the strong 
coupling approximation become exact over the entire coupling regime.

We insert a color dipole state into the vacuum described by the 
ground state wave functional.
The construction of the hadronic dipole is guided by the principles 
of the strong coupling 
approximation, i.e. the Schwinger string connecting the quark and the 
antiquark is chosen to follow the minimal transversal path in between 
the quark and the antiquark. The lattice naturally defines the hadronic 
state in configuration space. This is in contrast to other light cone 
lattice approaches like the transverse lattice approach \cite{Dalley:2003aj} 
where so-called ``fat links'' are quantised canonically and as such have 
an explicit formulation in the momentum representation which is more
natural for the computation of structure functions. In our approach,
we project the configuration space states explicitly on states with 
definite momenta in such a way, that each of the links forming the 
Schwinger string has its own momentum. Only the total momentum of the 
Schwinger string is constrained by the total hadron momentum (minus
the quark and antiquark momenta). Since our quarks are not dynamical, 
we cannot obtain their momenta from within our calculation. Therefore 
we take them from experiment. We use the average total string momentum 
obtained from the \emph{MRST} (2002) NLO parameterization of the gluon 
distribution function at $Q^2=1.5\,\mathrm{GeV}^2$ as an input.

The so obtained gluon distribution function obeys a recursive equation 
which relates the gluon distribution function with $n$ transversal 
links to the gluon distribution function with $n-1$ transversal links. 
If one interprets the increase of link constituents at a fixed size of 
the dipole as an increase in the resolution of the probe, this recursion 
relation is the non-perturbative counterpart of the DGLAP equation. 
Indeed, with increasing number of transversal links, the gluon distribution 
function grows at small $x_B$ due to the fact that the available total 
gluonic momentum has to be distributed among more and more constituents. 
A string-splitting function represents the probability to find a string 
containing $n-1$ transversal links inside of a string with $n$ transversal 
links. 

Our results calculated for the QCD-coupling $\lambda=10=4/g^4$ roughly 
correlate with a transverse lattice spacing of $a_{\perp}=0.5-0.7 \mathrm{fm}$. 
They can be compared with phenomenological parton distributions, if we choose 
the number of links appropriately for the proton ($n=4$). The calculated 
low $x$ gluon structure function shows a behavior similar to the 
\emph{MRST}-parametrization, once we fix the mean $\langle x_B\rangle$ in 
accord with these. Unfortunately, due to the lack of quark dynamics the mean 
gluon momentum $\langle x_B \rangle$ itself is out of reach.

The model presented here also shows that $x_B\, g(x_B) $ for the gluon at small 
$x$ becomes proportional to the hadronic size $ R_\perp^2$. This coincides 
with the empirical soft Pomeron behavior of hadronic cross sections. Both, 
the evolution of the structure function with increasing resolution $Q^2$ 
and/or with decreasing $x_B$ need a more sophisticated ground state wave 
functional (respecting scaling with the lattice spacing) and numerical 
simulations (corresponding to the inevitably non-locale action).
In previous work an improved wave functional has been 
proposed \cite{gruenewald_thesis} which can be used for structure function 
calculations, once it has passed the scaling tests in the light cone limit
$\eta\rightarrow 0$.

\begin{acknowledgments}
D.~G. acknowledges funding by the European Union 
project EU RII3-CT-2004-506078 and the GSI Darmstadt.
\end{acknowledgments}

\newpage
\begin{appendix}

\section{Relation of $g_n$ to $g_{n-1}$}
\label{app:Relgn2g1}

The dipole matrix element is related to $F_n(p_-^S)$ which is given by
\be
F_n(p_-^S)=\Big(\prod_{j=1}^n\,\sum_{y_j^-,y_j^\prime{}^-}\,\Big(\onehalf\Tr{U_{-k}}\Big)^{|y_j^--y_j^\prime{}^-|}\Big)\, \rho_{n}[p_-^S;y_j^-,y_j^\prime{}^-]\,.
\ee
With this definition, we
first prove the following relation for $F_n(p_-^S)$:
\be
F_n(p_-^S)=\sum_{p_-^{S\,\prime}=0}^{p_-^S} F_1(p_-^{S\,\prime})\,F_{n-1}(p_-^S-p_-^{S\,\prime})\,.
\label{eqn:NormRelApp}
\ee
One can use the definition of the density matrix and can insert a unity in form of two additional momentum summations with appropriate Kronecker deltas in order to obtain
\bea
F_n(p_-^S)&=&\sum_{p_-,p_-^\prime}\bigg[\sum_{y_1^-,y_1^\prime{}^-}\,\Big(\onehalf\Tr{U_{-k}}\Big)^{|y_1^--y_1^\prime{}^-|} 
\sum_{l_-^{1},l_-^{1\,\prime}}\,\delta(p_--l_-^{1})\,\delta(p_-^\prime-l_-^{1\,\prime})
\bigg.\nn\\
& &~~~~\bigg. e^{-\mathrm{i}\, \big(l_-^{1}/2\,y_1^--l_-^{1\,\prime}/2\,y_1^\prime{}^-\big)}\bigg]
\bigg[
\Big(\prod_{j=2}^n\,\sum_{y_j^-,y_j^\prime{}^-}\,\Big(\onehalf\Tr{U_{-k}}\Big)^{|y_j^--y_j^\prime{}^-|}
\sum_{l_-^{j},l_-^{j\,\prime}}\Big)\bigg.\nn\\ 
& &~~~~\bigg.
\delta(p_-^S-p_--\sum_{j=2}^n\,l_-^{j})
\,\delta(p_-^S-p_-^\prime-\sum_{j=2}^n\,l_-^{j\,\prime})\,e^{-\mathrm{i}\,\sum\limits_{j=2}^n \big(l_-^{j}/2\,y_j^--l_-^{j\,\prime}/2\,y_j^\prime{}^-\big)}\bigg]
\,.
\label{eqn:NormalizationEval}
\eea 
The first square bracket in the above equation is evaluated easily by performing a variable transformation and is given by
\bea
\lefteqn{
\sum_{y_1^-,y_1^\prime{}^-}\,s_1^{|y_1^--y_1^\prime{}^-|} 
\sum_{l_-^{1},l_-^{1\,\prime}}\,\delta(p_--l_-^{1})\,\delta(p_-^\prime-l_-^{1\,\prime})
\,e^{-\mathrm{i}\, \big(l_-^{1}/2\,y_1^--l_-^{1\,\prime}/2\,y_1^\prime{}^-\big)}}\nn\\
&=& 2\,N\,\delta(p_--p_-^\prime)\,\sum_{y_1^-}\,s_1^{|y_1^-|}\,e^{-\mathrm{i}\, p_-/2\,y_1^-}\nn\\
&=& \delta(p_--p_-^\prime)\,\sum_{y_1^-,y_1^\prime{}^-}\,s_1^{|y_1^--y_1^\prime{}^-|} 
\sum_{l_-^{1},l_-^{1\,\prime}}\,\delta(p_--l_-^{1})\,\delta(p_--l_-^{1\,\prime})
e^{-\mathrm{i}\, \big(l_-^{1}/2\,y_1^--l_-^{1\,\prime}/2\,y_1^\prime{}^-\big)}\nn\\
&=& \delta(p_--p_-^\prime)\,F_1(p_-)\,.
\eea
If one inserts this result into \eqnref{eqn:NormalizationEval}, one can perform the $p_-^\prime$ summation. After the evaluation of $\delta(p_--p_-^\prime)$, one can identify the second square bracket with $F_{n-1}(p_-^S-p_-^{S\,\prime})$ and \eqnref{eqn:NormRelApp} is proven. 

By using the same method to split of the contribution of a single link
from the entire matrix element, one can show that the following relation holds for the
gluon distribution function of a dipole containing $n$ transversal links:
\be
g_n(p_-^g;p_-^S)=\frac{n}{F_n(p_-^S)}\sum_{p_-^{S\,\prime}=0}^{p_-^S}\Big(F_1(p_-^{S\,\prime})\,g_1(p_-^g;p_-^{S\,\prime})\Big)\,F_{n-1}(p_-^S-p_-^{S\,\prime}) \, .
\label{eqn:RecRelgng1}
\ee
The factor of $n$ in front of the sum is due to the fact that the correlation function 
successively acts on each of the transversal links assembling the 
$n$-link dipole state when summed over the entire transversal lattice. The factor 
$1/F_{n}(p_-^S)$ ensures the correct 
normalization of the $n$-link dipole state. The sum goes over $F_1(p_-^{S\,\prime})\,g_1(p_-^g;p_-^{S\,\prime})$ times $F_{n-1}(p_-^S-p_-^{S\,\prime})$.

By using \eqnref{eqn:RecRelgng1}, one can rewrite the recursion relation in
terms of $g_{n-1}$ in order to obtain
\be
g_n(p_-^g;p_-^S)=\left(\frac{n}{n-1}\right) \frac{1}{F_n(p_-^S)}\sum_{p_-^{S\,\prime}=0}^{p_-^S} g_{n-1}(p_-^g;p_-^{S\,\prime})\,
F_{n-1}(p_-^{S\,\prime})\,F_1(p_-^S-p_-^{S\,\prime}) \, .
\ee

\end{appendix}

\end{document}